\documentstyle[jkas35,epsf]{article}

\runningauthor{TANIGUCHI et al.}
\runningtitle{HIGH-REDSHIFT LYMAN$\alpha$ EMITTERS}
\beginpage{1}
\endpage{23}

\begin{document}

\title{Lyman$\alpha$ Emitters beyond Redshift 5:
The Dawn of Galaxy Formation}

\author{Yoshiaki Taniguchi, Yasuhiro Shioya, Masaru Ajiki,
Shinobu S. Fujita, Tohru Nagao, \& Takashi Murayama}

\address{Astronomical Institute, Graduate School of Science,
         Tohoku University, Aramaki, Aoba, Sendai 980-8578, Japan \\
         {\it E-mail: tani@terra.astr.tohoku.ac.jp}}

\address{\normalsize{\it (Received June 01, 2003; Accepted June 1, 2003)}}

\abstract{
The 8m class telescopes in the ground-based optical astronomy 
together with help from the ultra-sharp eye of the Hubble Space Telescope
have enabled us to observe forming galaxies beyond redshift $z=5$.
In particular, more than twenty Ly$\alpha$-emitting galaxies
have already been found at $z > 5$. These findings provide us with useful hints
to investigate how galaxies formed and then evolved in
the early universe. Further, detailed analysis of Ly$\alpha$
emission line profiles are useful in exploring the nature
of the intergalactic medium because the trailing edge of
cosmic reionization could be close to $z \sim 6$ -- 7, at which 
forming galaxies have been found recently.
We also discuss the importance of superwinds from forming
galaxies at high redshift, which has an intimate relationship
between galaxies and the intergalactic medium.
We then give a review of early cosmic star formation history 
based on recent progress in searching for Ly$\alpha$-emitting 
young galaxies beyond redshift 5. 
}

\maketitle

\section{INTRODUCTION}

Formation and evolution of galaxies have been intensively discussed
in this decade. In particular, the great success of the Hubble Deep 
Field (HDF) project (Williams et al. 1996; 2000) has provided us with
very fruitful and firm observational results on faint galaxies
at high redshift through  systematic analyses of their broad-band optical
and near-infrared (NIR) color properties (e.g., Madau et al. 1996; Lanzetta
et al. 1996; Yahata et al. 2000). Follow-up optical and NIR spectroscopy
of faint galaxies found in the HDFs resulted in the discovery of many galaxies
beyond $z=3$; so-called Lyman break galaxies (LBGs)
(Steidel et al. 1996b; Lowenthal et al. 1997; Cohen et al. 2000;
Dawson et al. 2001). 
In addition to these HDF observations, several excellent
deep surveys in blank fields also contributed to the understanding
of early evolution of galaxies (Cowie et al. 1996; Steidel et al. 1996a,
1999, 2000; Pettini et al. 2001; Shapley et al. 2001, 2003).
This progress has also been supported by ground-based 8-10m class
telescopes such as Keck 10m telescopes, VLT, and Subaru Telescope.

The first discovery of a galaxy beyond $z=5$ was reported by Dey et al.
(1998); 0140+326 RD1 at $z=5.34$. This identification was made by detecting 
the redshifted Ly$\alpha$ emission line ($\lambda_{\rm rest} =1216$ \AA).
Since the classical paper by Partridge \& Peebles (1967), many challenges have
been actually made to search for Ly$\alpha$ emission from very young galaxies
at high redshift. Although most of those early attempts ended in negative results
before the mid 1990s (e.g., Pritchet 1994; Thompson et al. 1995),
recent advance in optical
spectroscopic capability with the 8-10m class telescopes
has enabled us to identify such very high-$z$ Ly$\alpha$ emitters (LAEs).
Following the success of Weymann et al. (1998), more than twenty
such very high-$z$ galaxies have been  found to date (see section III).

These galaxies are considered to be located at the trailing edge of the cosmic
reionization era and thus we are just facing close to the door of the dark age
of the universe (Becker et al. 2001; Djorgovski et al. 2001; 
see for a review, Loeb \& Barkana 2001; Barkana \& Loeb 2001). 
Accordingly, we human beings are now capable of investigating the dawn of
galaxy formation which occurred at $z \gtrsim$ 6 -- 7.
In this review, we focus our attention to the recent success of deep 
surveys of Ly$\alpha$-emitting forming galaxies and discuss the
cosmic star formation history, and some implications on superwind 
activities in such forming galaxies.
See also the excellent reviews on 
this research field given by Thommes (1999), Hu, Cowie, \& McMahon (1999),
and Stern \& Spinrad (1999).

We adopt a flat universe 
with $\Omega_{\rm matter} = 0.3$,
$\Omega_{\Lambda} = 0.7$, and $h=0.7$ where $h = H_0/($100 km s$^{-1}$
Mpc$^{-1}$) throughout this review.

\section{LYMAN$\alpha$ EMITTERS AT $2 < z < 5$}

\subsection{Introduction}

It has long been thought that forming galaxies are
strong emission-line objects because massive stars formed
in the initial phase could ionize the surrounding gas clouds.
Since such forming galaxies are located at high redshift,
the brightest emission line in the optical window could be
the hydrogen Ly$\alpha$ line (e.g., Partridge \& Peebles 1967;
Meier 1976). This prediction led many astronomers to search for
high-$z$ LAEs (see for a review  Pritchet 1994).
Custom narrowband filters (the band pass is $\simeq$ 100 \AA)
have often often used to search for the Ly$\alpha$ emission line.
The central wavelength is of course tuned to the wavelength 
of the redshifted Ly$\alpha$.

Prior discussing to the discovery of Ly$\alpha$ emitters (LAEs) beyond $z=5$,
we give a brief review of Ly$\alpha$ emitters between $z=2$ and $z=5$.
In this section, we also give a summary of survey methods that have
been used to find LAEs (see also Thommes 1999; Stern \& Spinrad 1999).

\subsection{Guided Surveys}

\subsubsection{Optical Counterparts of Quasar Absorption Line Systems}

Since the discovery in 1968 (Burbidge et al. 1968;
Bahcall et al. 1968),
several populations of absorption systems have been found in quasar spectra.
Such quasar absorption systems are classified as follows.
1) Damped Ly$\alpha$ absorption systems (DLAs) with
$N$(H {\sc i}) $\gtrsim 2 \times 10^{20}$ cm$^{-2}$,
2) Lyman limit absorption systems (LLS) with
$N$(H {\sc i}) $\gtrsim 1.6 \times 10^{17}$ cm$^{-2}$,
3) Sub Lyman limit absorption systems (LLS) with
$N$(H {\sc i}) $\sim  10^{16}$ cm$^{-2}$,
4) Ly$\alpha$ forests with
$N$(H {\sc i}) $\sim  10^{12 - 15}$ cm$^{-2}$, and
5) Metal-line absorption systems probed either by
low-ionization absorption lines such as Mg {\sc ii} or by
high-ionization absorption lines such as C {\sc iv};
Mg {\sc ii} absorbers have $N$(H {\sc i}) $\gtrsim 10^{17}$ cm$^{-2}$ while
C {\sc iv} absorbers have $N$(H {\sc i}) $\gtrsim 10^{16}$ cm$^{-2}$.
These systems have been utilized to investigate both protogalactic disks/halos
and the intergalactic matter at high redshift
(e.g., Weymann et al. 1981; Tytler 1982; Wolfe et al. 1995).
Among the absorption system  population,
much attention has been paid to Damped Ly$\alpha$
absorption systems (DLAs)
because their observed H {\sc i} column densities, $> 2 \times 10^{20}$ cm$^{-2}$,
are comparable to those of galactic disks and thus
they are often considered to be progenitors of the present-day
disk galaxies. 
Possible origins of both DLAs and LLSs have been considered as;
1) Protogalactic disks/halos,
2) merging protogalactic gas clumps in CDM cosmologies,
3) gas-rich dwarf galaxies,
4) collapsing halos with merging clouds,
and 5) low surface brightness galaxies.
Therefore, it is important to
identify counterparts of DLAs and LLSs at high redshift.

Although early searches mostly failed, the first discovery of
a LAE beyond $z=2$ was reported by Hunstead et al. in 1990.
They found a Ly$\alpha$ emitting object as an optical counterpart of  
the Damped Ly$\alpha$ absorption (DLA) system at $z_{\rm abs} =
2.465$ which was detected toward the line of sight to Q0836+113
at $z_{\rm em} = 2.67$. This LAE is small ($\lesssim$ 10 kpc)
and has a very narrow line width (FWHM $<$ 60 km s$^{-1}$). 
The inferred star formation rate (SFR\footnote{In this review,
we use the relation $SFR = 9.1 \times 10^{-43} L({\rm Ly}\alpha) ~ M_\odot
{\rm yr}^{-1}$ (Kennicutt 1998; Brocklehurst 1971).})
is also modest, $SFR \sim 1 M_\odot$ yr$^{-1}$.
Therefore, they suggested that this LAE is a gas-rich H {\sc ii} galaxy.

However, this observation was not confirmed by later observations
(Wolfe et al. 1992; Lowenthal et al. 1995).
Nevertheless, their work encouraged researchers to carry out similar searches
for LAE counterparts of DLAs at high redshift because this method provided
a new type of guided search for high-$z$ LAEs. 
As summarized in Table 1, up to now, LAE counterparts
of the seven DLA systems have been confirmed (e.g., Djorgovski et al. 1996)
in the redshift range between $z=1.9$ and $z=3.4$.

The most distant LAE found so far as a counterpart of DLA
is DLA0000$-$2619 at $z=3.39$ (Macchetto et al. 1993; Giavalisco et al. 1994).
The second most distant known is DLA 2233+131 at $z=3.15$ (Djorgovski et al. 1996).
The latter object appears to be a massive disk galaxy with $M \sim 10^{11} M_\odot$.
This reinforces the importance of the optical identification of
LAE counterparts of DLAs at high redshift even though the success rate
is not so high.
We hope that such surveys will be applied to higher-$z$ DLAs found in 
very high-$z$ SDSS quasars (e.g., Fan et al. 2000, 2001, 2003)

One interesting problem in the optical counterparts is that 
some of them have large impact parameters; e.g., a few hundreds
kpc from the line of sight toward a quasar. If they are really
absorbing agents, their halo size is much larger than for typical galaxies
[see also section IV (f)].

%-----------------------------------------------------
%    Table 1
%-----------------------------------------------------
{
\scriptsize
\begin{deluxetable}{lccc}
\tablenum{1}
\tablewidth{6.5in}
\tablecaption{%
A summary of the optical counterpart searches for DLAs
}

\tablehead{
\colhead{QSO} & \colhead{$z_{\rm em}$} & \colhead{$z_{\rm abs}$} &
\colhead{Ref\tablenotemark{a}} \nl
%\cline{5-6}
 &  &  & \colhead{(cm$^{-2}$)} \nl
}
\startdata
Q0000-2619           & 4.36  & 3.3902 & 1, 2 \nl
Q0100+1300 (PHL 957) & 2.681 & 2.309  & 3    \nl
Q0151+048A           & 1.922 & 1.9342 & 4    \nl
PKS 0528-250         & 2.77  & 2.811  & 5, 6 \nl
Q2059-360            & 3.097 & 3.0831 & 7    \nl
Q2139-4434           & 3.23  & 2.380  & 8, 9 \nl
Q2233+131            & 3.295 & 3.150  & 10   \nl
\enddata
\tablenotetext{a}{1. Macchetto et al. 1993, 2. Giavalisco et al. 1994,
3. Lowenthal et al. 1991, 4. M\"oller et al. 1998, 5. M\"oller \& Warren 1993,
6. Warren \& M\"oller 1996, 7. Leibundgut \& Robertson 1999, 8. Francis et al.
1996, 9. Francis et al. 1997, and 10. Djorgovski et al. 1996.}
\end{deluxetable}
}

\subsubsection{Objects Associated with High-$z$ Active Galactic Nuclei}

Another type of guided search for high-$z$ LAEs are searches for 
LAEs at a particular redshift where the presence of a certain high-$z$
source is already known; e.g., active galactic nuclei (AGNs)
such as quasars and radio galaxies.
Note that the association does not always mean physical companion 
galaxies. Rather than this, such searches are intended to find 
galaxies associated with large-scale structure (e.g., a protocluster
region) to which the target AGN also belongs. 

The first successful result was reported by Djorgovski et al. (1985)
who found a Ly$\alpha$-emitting companion galaxy ($z=3.22$) associated
with a radio-loud quasar PKS 1614+051 at $z=3.21$ (see also Djorgovski
et al. 1987). 
Steidel et al. (1991) also found a Ly$\alpha$-emitting companion 
($z=2.758$) associated with a radio-quiet quasar Q1548+0917
at $z=2.749$. 

Hu et al. (1992) conducted a systematic search
for Ly$\alpha$ companion galaxies for a sample of 17 quasars.
They found companion candidates for three out of 10 radio-loud
quasars while found no candidates were found for seven radio-quiet ones. 
Later, Hu et al. (1996) found a Ly$\alpha$ companion of BR 1202$-$0725
at $z=4.69$. Its Ly$\alpha$ luminosity leads to the star formation
rate of several $M_\odot$ yr$^{-1}$ (see also Ohta et al. 2000).

One problem in these surveys is that photoionization sources
in such close companion galaxies are relatively uncertain
because the intense radiation from the quasar itself affects 
the ionization of gas clouds in companion galaxies.
Indeed, the companion galaxy to PKS 1614+051 shows high-ionization
emission lines such as C {\sc iv} in its UV spectrum.

More recently, Hu \& McMahon (1996)
searched for LAEs in the field of quasar BR2237$-$0607 at $z=4.55$
using a narrowband filter centered on the quasar's redshifted Ly$\alpha$
emission using the UH 2.2 m telescope. They found 10 LAE candidates in the
narrowband image and confirmed spectroscopically two LAEs at  $z=4.55$.
Since these two objects are far from the quasar, it is expected that
the ionization is attributed to photoionization by massive stars.
Keel et al. (1999) made a similar survey in the field of the radio
galaxy 53W 002 at $z=2.4$ and found $\approx$ 20 LAEs at the same redshift.
A summary of the above surveys is given in Table 2.

%=======================================================
%      Table 2
%=======================================================
{
\scriptsize
\begin{deluxetable}{lccccccc}
\tablenum{2}
\tablewidth{6.5in}
\tablecaption{%
A summary of the guided surveys for Ly$\alpha$-emitter at $2 < z < 5$ 
}

\tablehead{
  \colhead{Survey\tablenotemark{a}} &
  \colhead{Field\tablenotemark{b}} &
  \colhead{$z_{\rm c}$\tablenotemark{c}} &
  \colhead{$(z_{\rm min}, z_{\rm max})$\tablenotemark{d}} &
  \colhead{$V$\tablenotemark{e}} &
  \colhead{$EW_{\rm lim}({\rm Ly}\alpha)$\tablenotemark{f}} &
  \colhead{$N({\rm Ly}\alpha)$\tablenotemark{g}} &
  \colhead{$n({\rm Ly}\alpha)$\tablenotemark{h}}
}
\startdata
HM96 & BR2237$-$0607 & 4.55 & (4.52, 4.56) & 2304 & \nodata & 2 & $8.7 \times 10^{-4}$ \nl 
K99 & 53W002 & 2.4 & (2.32, 2.45) & 85338 & 92 & 19 & $2.2 \times 10^{-4}$ \nl
K99 & 53W002E & 2.55 & (2.49, 2.61) & 78588 & 291 & 1 & $1.3 \times 10^{-5}$ \nl
K99 & 53W002N & 2.55 & (2.49, 2.61) & 78588 & 155 & 1 & $1.3 \times 10^{-5}$ \nl
K99 & 53W002NE & 2.55 & (2.49, 2.61) & 78588 & 184 & 4 & $5.1 \times 10^{-5}$ \nl
\enddata
\tablenotetext{a}{HM96 = Hu \& McMahon (1996), and K99 = Keel et al. (1999).}
\tablenotetext{b}{The name of the targeted field.}
\tablenotetext{c}{The central redshift corresponding to the central
     wavelength of the narrow-band filter ($\lambda_{\rm c}$).}
\tablenotetext{d}{The minimum and maximum redshift covered by the
      narrow-band filter.}
\tablenotetext{e}{The co-moving volume covered by the survey in
      units of $h_{0.7}^{-3}$  Mpc$^3$ with $\Omega_{\rm matter} = 0.3$
      and $\Omega_\Lambda = 0.7$.}
\tablenotetext{f}{The smallest equivalent width of the Ly$\alpha$ emission
      detected in the survey in units of \AA ~ in the observed frame.}
\tablenotetext{g}{The number of Ly$\alpha$ emitters found in the survey.}
\tablenotetext{h}{The number density of Ly$\alpha$ emitters found in the survey
      in units of $h_{0.7}^3$ Mpc$^{-3}$.}
\tablenotetext{i}{The LBG (Lyman break galaxies) spike region.}
\tablenotetext{j}{La Palma field (Mendez et al. 1997) in the Virgo cluster.}
\end{deluxetable}
}

\subsubsection{Powerful Radio Galaxies}

Powerful radio galaxies at high redshift have been providing nice tools
that can be used 
to investigate observational properties of young galaxies hosting
radio sources. However, this issue is out of the scope of this review.
We recommend readers to visit 
the following nice reviews (McCarthy 1993; Stern \& Spinrad 1999).

\subsubsection{Systematic Serendipitous Searches for LAEs}

Serendipitous discoveries of LAEs at high redshift have
sometimes been reported in this decade (e.g., Franx et al. 1997;
Dey et al. 1998; Dawson et al. 2001; Lehnert \& Bremer 2003). 
Such objects were really
found serendipitously. However, as a survey technique, there is 
a {\it systematic} serendipitous search for high-$z$ galaxies. 
In this technique, a certain spectroscopic database is used.
Since most spectroscopic observations could be made to investigate
some target objects, we introduce this technique in the category of
guided surveys.

Thompson \& Djorgovski (1995) reported their ^^ ^^ systematic"
serendipitous search for LAEs using a database of optical, long-slit
spectroscopy obtained with both the Double Spectrograph and 
4-shooter instruments  on the Hale 5m telescope. The database
consists of 421 independent spectra covering 14.97 arcmin$^2$;
note that 1$\sigma$ limiting flux is $\sim 1 \times 10^{-17}$
ergs s$^{-1}$ cm$^{-2}$.
They found 65 emission-line objects among which  
are two quasars at $z=1.67$ and $z=1.19$. 

Using much deeper spectroscopic archival data obtained by Keck
telescope, Manning et al. (2002) made a similar search for 
high-$z$ LAEs.  They found two LAE candidates at $z=3.02$ and
$z=3.36$. Since these objects are close to the Lyman break galaxy
SSA22-C17 at $z=3.30$, they may be objects associated with 
the overdensity region identified by Steidel et al. (2000).

Our group actually has the experience of the serendipitous discovery of 
a high-$z$ object (a BAL quasar at $z=3.1$) in our long-slit
spectrum  obtained with ESI on Keck II telescope. 
Spectroscopic archival data must be checked carefully by
anyone who is interested in serendipitous discovery of 
high-$z$ objects. 
This may provide an example of the importance of the Virtual
Observatory (see for a recent review, Brunner et al. 2002).

\subsection{Blank-Field Surveys}

One of main purposes of deep surveys is to understand how
ordinary galaxies like the Milky Way formed and  evolved.
For this purpose, it is important to observe typical fields in which
there are no known unusual object such as luminous AGNs and massive
clusters of galaxies; i.e.,
blank fields. Famous blank fields are the Hubble Deep Fields
(HDFs: Williams et al. 1996, 2000)
and the SSA (small selected area) fields selected by the L. L. Cowie
team at IfA, UH (Cowie et al. 1996 and references therein).

\subsubsection{Spectroscopic Follow-up of Objects found in
Black-Field Surveys}

Since the early 1990s, the so-called Lyman break method has often been
used to identify high-$z$ galaxies (Steidel \& Hamilton 1992,
1993; Steidel et al. 1995, 1998, 1999; Lanzetta et al. 1996; Madau et al. 1996;
Yahata et al. 2000; Iwata et al. 2003; Yan et al. 2002).
Since such surveys provided nice samples of
LBGs at $z>3$, follow-up spectroscopy has been
conducted intensively and a large number of high-$z$ LBGs 
have already been identified (Steidel et al. 1996a, 1996b; Lowenthal et al. 1997;
Shapley et al. 2003 and references therein).

Another important discovery came from a spectroscopic survey for 
galaxies toward a foreground cluster of galaxies at $z=0.37$;
MS 1512-cB58 at $z=2.73$ (Yee et al. 1996). This object is
apparently bright for such a high-$z$ LBG ($V=20.6$) and thus
it provides a nice laboratory for investigations of young galaxies
(Pettini et al. 2000 and references therein). 

\subsubsection{Searches with a Narrowband Filter}

Deep surveys for LAEs beyond $z=2$ with a narrowband filter were 
already made in the early 1990s. 
Thompson et al. (1995) used a Fabry-Perot interferometer to
search for LAEs with $z = 2.78$ - 4.89. Although their 1$\sigma$
limiting flux was down to $\sim 1.5 \times 10^{-17}$ ergs s$^{-1}$ cm$^{-2}$
over 0.05 deg$^2$ and $\sim 8.5 \times 10^{-17}$ ergs s$^{-1}$ cm$^{-2}$
over 0.63 deg$^2$, they found no LAE candidates.
 
The first attempt by Cowie \& Hu (1998) with the Keck 10m telescope revealed
the presence of five Ly$\alpha$ emitters in  HDF-N (N=North) and
seven in SSA22 at $z \approx 3.4$. The star formation rates of the LAEs are
several $M_\odot$ yr$^{-1}$ and the star formation rate density is 
$\rho_{\rm SFR} \sim 0.01 M_\odot$ Mpc$^{-3}$ yr$^{-1}$.  
Subsequently, Steidel et al. (2000) also succeeded in
finding 77 LAE candidates at $z=3.1$ in a blank field around SSA22.
Furthermore, Kudritzki et al. (2000) found nine LAEs at $z=3.1$ 
during the course of their narrow-band imaging survey aimed at
looking for intracluster planetary nebulae in the Virgo cluster
(Mendez et al. 1997).

Wide-field deep surveys were conducted by using 
the Suprime-Cam  mounted at the prime focus of
the 8.2 m Subaru telescope at Mauna Kea; note that 
this camera covers a sky area of 34$^\prime \times 27^\prime$.
Using a narrowband filter centered at 7110 \AA ~ (NB711),
Ouchi et al. (2003) found 87 LAE candidates at $z=4.86$ in the Subaru
Deep Field (SDF: see for the SDF, Maihara et al. 2001).

Fujita et al. (2003) used an intermediate band filter
with a resolution of $R=23$ centered at 5736 \AA ~ (IA574) on
Suprime-Cam in the Subaru XMM Deep Field (SXDF)
and found six LAE candidates at $z \approx 3.7$. This filter is one of 
a set of special purpose filters designed for Suprime-Cam
(Hayashino et al. 2000; Taniguchi 2001).

Similar wide-field deep surveys were also conducted by using
the CCD mosaic camera at the Kitt Peak National Observatory's 4m
Mayall telescope; i.e., the LALA survey [Rhoads et al. 2000;
Malhotra \& Rhoads 2002; see also section VI (b)].
This camera covers a sky area of $36^\prime \times 36^\prime$
and thus this is very suitable for all optical surveys.
Rhoads et al. (2000) used five narrowband filters centered at
6559 \AA, 6611 \AA, 6650 \AA, 6682 \AA, and 6730 \AA, each of
which has a passband of $\approx$ 80 \AA. This filter set
made it possible to search for LAEs with $4.37 < z < 4.57$;
as well as for their search for LAEs at $z \approx 5.7$, see next section.

We give a summary  of 
the blank-field surveys for Ly$\alpha$-emitters at $2 < z < 5$
in Table 3.

%=======================================================
%      Table 3
%=======================================================
{
\scriptsize
\begin{deluxetable}{lccccccc}
\tablenum{3}
\tablewidth{6.5in}
\tablecaption{%
A summary of the blank-field surveys for Ly$\alpha$-emitter at $2 < z < 5$
}

\tablehead{
  \colhead{Survey\tablenotemark{a}} &
  \colhead{Field\tablenotemark{b}} &
  \colhead{$z_{\rm c}$\tablenotemark{c}} &
  \colhead{$(z_{\rm min}, z_{\rm max})$\tablenotemark{d}} &
  \colhead{$V$\tablenotemark{e}} &
  \colhead{$EW_{\rm lim}({\rm Ly}\alpha)$\tablenotemark{f}} &
  \colhead{$N({\rm Ly}\alpha)$\tablenotemark{g}} &
  \colhead{$n({\rm Ly}\alpha)$\tablenotemark{h}}
}
\startdata
CH98 & HDF   & 3.4 & (3.41, 3.47) & 5205 & 115 & 5 & $9.6 \times 10^{-4}$ \nl
CH98 & SSA22 & 3.4 & (3.41, 3.47) & 5205 & 90 & 7 & $1.3 \times 10^{-3}$  \nl
K99  & HU Aqr & 2.4 & (2.32, 2.45) & 85338 & 241 & 1 & $1.2 \times 10^{-5}$ \nl
K99  & NGC 6251 & 2.4 & (2.32, 2.45) & 85338 & \nodata & 0 & 0 \nl
S00  & LBGS\tablenotemark{i} & 3.09 & (3.07, 3.12) & 16741 & 80 & 72
        & $4.3 \times 10^{-3}$ \nl
K00  & Virgo\tablenotemark{j} & 3.14 & (3.12, 3.15) & 6020 & \nodata &
         8 & $1.3 \times 10^{-3}$ \nl
R00  & LALA\tablenotemark{k} & 4.4 & (4.37, 4.43) & 212000 &
        80  & 225 & $1.1 \times 10^{-3}$  \nl
M02  & LALA\tablenotemark{l} & 4.4 & (4.37, 4.57) & 740000 &
        80  & 157 & $2.1 \times 10^{-4}$  \nl
O03  & SDF    & 4.9 & (4.83, 4.89) & 89600 & 82 & 62\tablenotemark{m} 
          & $6.9 \times 10^{-4}$  \nl
F03  & Subaru/XMM & 3.72 & (3.60, 3.83) &  93952 & 254 & 6
           &  $6.4 \times 10^{-5}$ \nl
\enddata
\tablenotetext{a}{CH98 = Cowie \& Hu (1998), K99 = Keel et al. (1999),
     S00 = Steidel et al. (2000), K00 = Kudritzki et al. (2000), R00 =
     Rhoads et al. (2000), M02 = Malhotra \& Rhoads (2002), O03 =
     Ouchi et al. (2003), and F03 = Fujita et al. (2003).}
\tablenotetext{b}{The name of the targeted field.}
\tablenotetext{c}{The central redshift corresponding to the central
     wavelength of the narrow-band filter ($\lambda_{\rm c}$).}
\tablenotetext{d}{The minimum and maximum redshift covered by the
      narrow-band filter.}
\tablenotetext{e}{The co-moving volume covered by the survey in
      units of $h_{0.7}^{-3}$  Mpc$^3$ with $\Omega_{\rm matter} = 0.3$
      and $\Omega_\Lambda = 0.7$.}
\tablenotetext{f}{The smallest equivalent width of the Ly$\alpha$ emission
      detected in the survey in units of \AA ~ in the observed frame.}
\tablenotetext{g}{The number of Ly$\alpha$ emitters found in the survey.}
\tablenotetext{h}{The number density of Ly$\alpha$ emitters found in the survey
      in units of $h_{0.7}^3$ Mpc$^{-3}$.}
\tablenotetext{i}{The LBG (Lyman break galaxies) spike region.}
\tablenotetext{j}{La Palma field (Mendez et al. 1997) in the Virgo cluster.}
\tablenotetext{k}{We refer only to their survey results with the NB6599 filter.}
\tablenotetext{l}{Five NB filters centered at 6559 \AA, 661 \AA, 6650 \AA, 
                  6692 \AA, and 6730 \AA are used.}
\tablenotetext{n}{They found 87 candidates in total. However, they secured that
      62 sources are much more reliable because of their $R-i^\prime$ colors.}
\end{deluxetable}
}

\subsection{Gravitationally-Amplified-Field Surveys for 
Ly$\alpha$ Emitters at $2 < z < 5$}

Basically, there is no survey for LAEs at $2 < z < 5$
in a gravitationally-amplified field although this method
has been recently used to search for LAEs beyond $z=5$
(Ellis et al. 2001; Hu et al. 2002; see next section).
However, it is naturally understood that any help from 
gravitational lensing is highly useful to find 
distant LAEs because such LAEs are often very faint.
In order to demonstrate this, we introduce an epoch-making
serendipitous discovery of LAEs at $z=4.92$ by Franx et al.
in 1997.

In order to study the observational properties of 
member galaxies of the cluster Cl 1357+62
at $z = 0.33$, they obtained a large mosaic of multi-color
WFPC2 images of the cluster using HST. Franx et al. (1997)
found a red arc
in their image and then they obtained optical spectroscopy of this arc
using LRIS at Keck Observatory.
Surprisingly, they identified this arc as a LAE at $z=4.92$.
They further found that a companion galaxy is also a LAE
at nearly the same redshift (the velocity difference is only 
450 km s$^{-1}$). These objects were the most distant ones
up to 1997 and their redshifts were very close to $z=5$.
Their discovery surely encouraged
astronomers who are interested in forming galaxies in the early
universe. Indeed, a similar survey made by Frye et al. (2002)
resulted in the discovery of
eight LAEs at $3.7 < z < 5.2$ (seven are at $z <5$)
which were selected
from their red broad-band colors behind four massive clusters
of galaxies, Abell 1689, Abell 2219, Abell 2390, and AC 114. 

\section{LYMAN$\alpha$ EMITTERS BEYOND REDSHIFT 5}

\subsection{Introduction}

The recent advance provided by the 8-10m class telescopes
has also enabled us to carry out deep imaging  searches for star-forming galaxies
beyond redshift 5.
In particular, imaging surveys using narrow-passband filters have
proved to be an efficient way to find such galaxies (Hu \&
McMahon 1996; Cowie \& Hu 1998; Steidel et al. 2000; Kudritzki et
al. 2000; Rhoads et al. 2001; Rhoads \& Malhotra 2001;
Hu et al. 2002; Ajiki et al. 2002).  Indeed the most
distant Ly$\alpha$ emitter known to date, HCM 6A at $z=6.56$ has been
discovered by using this technique (Hu et al. 2002; see also Hu et al. 1998,
1999; Ajiki et al. 2002).

Other techniques also led to the discovery of
Ly$\alpha$ emitters beyond $z=5$ (Dey et al. 1998; Spinrad et al. 1998; Weymann
et al. 1998; Dawson et al. 2001, 2002; Ellis et al. 2001).
In this section, we give a summary of discoveries of LAEs beyond
$z=5$ based on various survey techniques.

\subsection{Spectroscopic Follow-up of Objects found in 
Deep Surveys}

The Lyman break method applied to HDF-N brought a number of 
candidate galaxies beyond $z=5$ (e.g, Lanzetta et al. 1996;
Madau et al. 1996). Subsequent spectroscopic observations 
revealed that HDF 4-473.0 is  really a LAE at $z=5.60$
(Weymann et al. 1998). Spinrad et al. (1998) also found two
galaxies (HDF 3-95.1 and HDF 3-95.2) at $z=5.34$.
Although these objects do not show strong Ly$\alpha$ emission,
they show a continuum break at $\lambda \simeq 7710$ \AA.
Two more galaxies beyond $z=5$ were found during a spectroscopic
survey of faint galaxies in the HDF-N flanking field;
F36218-1513 at $z=5.767$ and F36246-1511 at $z=5.631$ (Dawson et al. 2001). 
More recently, Lehnert \& Bremer (2003) found five  LAEs
beyond $z=5$ ($5.02 < z < 5.87$) from their high-$z$ LBG candidates. 

\subsection{Serendipitous Discoveries of LAEs beyond $z=5$}

The first galaxy beyond $z=5$ was found serendipitously; 0140+326
RD1 at $z=5.34$ (Dey et al. 1998). This object was found during 
their search for LBGs at $z \sim 4$.
Dawson et al. (2002) also found a LAE at $z=5.190$ serendipitously.
This object, J123649.2+621539, was found in a long-slit spectrum
of Keck/ESI during their spectroscopic search for faint galaxies
in the HDF-N NW flanking field. As summarized in Table 4, we know
25 confirmed (or most probable)
LAEs beyond $z=5$. Among them, two are serendipitously
discovered. This high discovery rate seems to be attributed to
the great observational capability in spectroscopy with 8-10m telescopes. 

\subsection{Surveys with a Narrowband Filter}

Recent surveys with a narrowband filter have been 
finding a number of LAEs beyond $z=5$. 
However, such surveys 
have the following two limitations.
The first limitation comes from strong OH airglow emission lines.
In Fig. 1, we show an
optical spectrum of OH airglow emission lines which is kindly
supplied by Alan Stockton (see also Stockton 1999). 
Although there can be seen several gaps at which there is little
strong OH emission, OH emission dominates at wavelengths longer than
7000 \AA. This prevents us from finding very faint galaxies at high redshift.
The well-defined gaps appear around $\lambda \approx$ 7110 \AA, 8160 \AA,
and 9210 \AA. These gaps enable us to search for LAEs at
$z \approx$ 4.8 (Rhoads et al. 2000; Ouchi et al. 2003), 5.7
(Hu et al. 1998; Ajiki et al. 2002, 2003; Taniguchi et al. 2003), and 6.6
(Hu et al. 2002; Kodaira et al. 2003).

%%%%%%%%%%%%%%%%%%%%%%%%
%   Fig. 1 (OH)
%%%%%%%%%%%%%%%%%%%%%%%%
\begin{figure}
\epsfysize=5.5cm \epsfbox{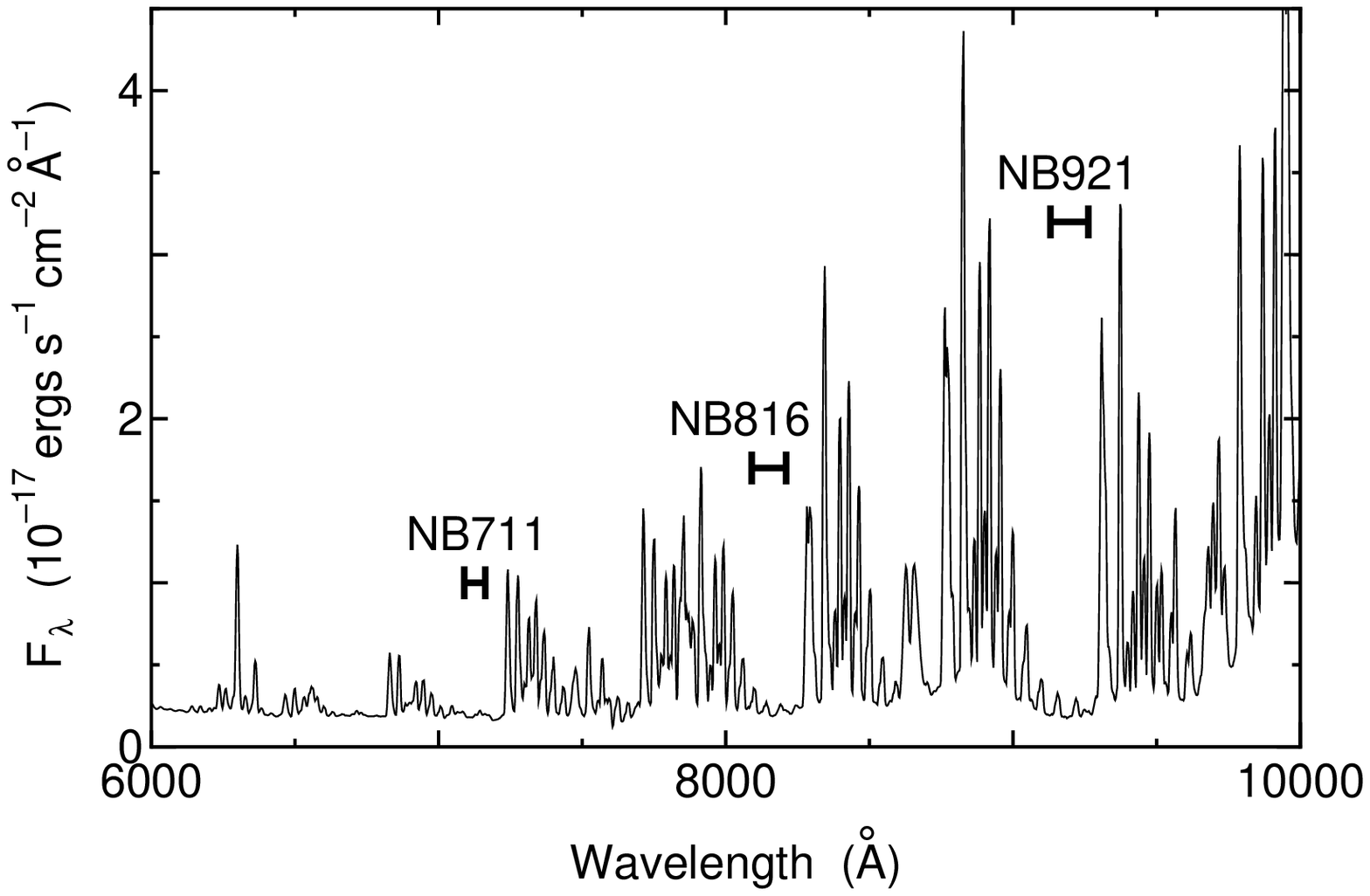}
%\plotone{fig1.eps}
%\vspace{10cm}
\caption[]{
Optical spectrum of OH airglow emission lines
taken from Stockton (1999).
The three narrowband filter bands, NB711, NB816, and NB921,
are shown by horizontal bars, all of that are available 
for the Suprime-Cam on the Subaru Telescope.
\label{fig1}
}
\end{figure}

The second limitation is that survey volumes are so small because of
narrower band widths (e.g., $\sim 100$ \AA). In order to gain survey volumes
and to reach faint limiting magnitudes,
we need wide-field CCD cameras on 8-10m class telescopes.
The Suprime-Cam  mounted at the prime focus of
the 8.2 m Subaru telescope at Mauna Kea provides a unique
opportunity for wide-field (a $34^\prime \times 27^\prime$ field of view),
narrowband imaging surveys for emission-line objects at high redshift.
The efficiency of this instrument is higher by a factor of 30 
than typical imagers mounted on the other 8-10m class telescopes.

Since the first
detection of a galaxy beyond $z=5$ (Dey et al. 1998), more than twenty such
galaxies have been found (Table 4). The most distant one known to date is HCM
6A whose redshift is 6.56 (Hu et al. 2002), farther than the most distant
quasar known, SDSSp J114816.64+525150.3 at $z=6.43$ (Fan et al. 2003).

Hu et al. (2002) made imaging surveys of LAEs at $z \approx 6.6$
for the following six fields using 
the LRIS on Keck I; three blank fields (HDF-N, SSA 17, and SSA 22) and
three cluster fields around Abell 370, Abell 851, and Abell 2390. 
Then they found several LAE candidates in their surveys
(Esther Hu, private communication) and confirmed
that HCM-6A found in the Abell 370 field
is a LAE at $z=6.56$.  This object shows a significant
continuum break at wavelengths shorter than the Ly$\alpha$ line peak 
and the relatively flat UV continuum emission. All these features 
are expected for a very young, forming galaxy at such a high redshift
(e.g., Meier 1976). The unlensed Ly$\alpha$ flux leads to a star
formation rate of $\sim 3 M_\odot$ yr$^{-1}$ with the adopted cosmology. 

Recently, some new survey results  with narrowband filters with the Suprime-Cam
on the 8.2 m Subaru telescope have appeared. 
First, our group made a deep survey for LAEs at $z=5.7$
(Ajiki et al. 2003) guided by one of the very high-$z$ SDSS quasars,
SDSSp J104433.04$-$012502.2 at $z=5.74$ (Fan et al. 2000).
They found approximately twenty LAE candidates and already confirmed
spectroscopically two LAEs at $z=5.69$ (Ajiki et al. 2002)
and $z=5.66$ (Taniguchi et al. 2003).
More recently, Kodaira et al. (2003) made a deep imaging survey for
LAEs at $z \approx 6.6$ using the NB921 filter with the Suprime-cam.
They succeeded in identifying LAEs at
$z=6.51$ and $z=6.58$ in the SDF [see also section VI (c)].
The latter LAE is more distant than HCM-6A at $z=6.56$.
These successful results reinforce that this method is highly useful in
searching for high-$z$ LAEs.

\subsection{Summary}

To conclude we give a summary in Table 4 of all the galaxies beyond
redshift 5 in the literature. New
discoveries of LAEs at $z>5$ recently reported (Rhoads et al.
2003; Lehnert \& Bremer 2003; Kodaira et al. 2003) are also
included in this table. However, we do not include a LAE
candidate at $z=6.5926$ found by Lehnert \& Bremer (2003)
because it seems necessary to obtain a high-resolution spectrum
for the confirmation.
We also do not include the object at $z=6.68$ reported by
Chen et al. (1999) because this redshift is not confirmed by later
observations (Stern et al. 2000b; Chen et al. 2000).
The radio galaxy, TN J0924$-$2201 at $z=5.19$ (van Breugel et al. 1999)
is not included too, because this source is an AGN.

As shown in Table 4, recent optical follow-up spectroscopy has used
high-resolution spectrographs like ESI (Sheinis et al. 2000)
 rather than low-resolution ones
like LRIS. The reason for this is that  there is a clear advantage in using
high-resolution ($R >$ 2000 - 3000) spectroscopy to extract Ly$\alpha$ from
the forest of OH airglow lines above $\sim$ 7000 \AA ~ beyond $z \sim$ 5
(see also Stern \& Spinrad 1999). 
High spectral
resolution will be particularly important in future searches for
subgalactic Ly$\alpha$ emitters at high redshift if the narrow Ly$\alpha$
profiles found in Abell 2218 (Ellis et al. 2001) and LAE J1044$-$0123
(Taniguchi et al. 2003) are typical of distant
emission-line objects (see section VI).

%-----------------------------------------------------
%    Table 4
%-----------------------------------------------------
{
\scriptsize
\begin{deluxetable}{rlcccccc}
\tablenum{4}
%\tabletypesize{\scriptsize}
\tablecaption{A summary of galaxies beyond redshift 5}
\tablewidth{0pt}
\tablehead{
\colhead{No.} &
\colhead{Name} &
\colhead{Redshift}  &
\colhead{$f$(Ly$\alpha$)\tablenotemark{a}}  &
\colhead{$L$(Ly$\alpha$)\tablenotemark{b}}  &
\colhead{Method\tablenotemark{c}}  &
\colhead{Sp. Mode\tablenotemark{d}} &
\colhead{Ref.\tablenotemark{e}}
}
\startdata
1  & SDF J132418.3+271455     & 6.578  & 2.1 & 10.0 & NB & L & 1 \\
2  & HCM 6A                   & 6.56   & 2.7 & 3.3 & NB & L & 2 \\
3  & SDF J132415.7+273058     & 6.541  & 1.1 & 5.6 & NB & L & 1 \\
4  & LAE @ 0226-04 Field      & 6.17   & 3.9 & 17  & SF & L & 3 \\
5  & BDF 1:19                 & 5.8696 & 0.31 & 1.2 & SF & L & 4 \\
6  & F36218-1513              & 5.767 & \nodata & \nodata & SF & L & 5 \\
7  & LALA J142546.76+352036.3 & 5.746 & 1.9 & 6.7 & NB & L & 6 \\
8  & BDF 1:10                 & 5.7441 & 2.4 & 8.7 & SF & L & 4 \\
9  & SSA22-HCM1               & 5.74  & 1.7 & 6.1 & NB & L & 7 \\
10 & LALA J142647.16+353612.6 & 5.700 & 3.9 & 14  & NB & L & 6 \\
11 & LAE J1044$-$0130         & 5.687 & 1.5 & 5.2 & NB & H & 8 \\
12 & LALA J142630.34+354022.5 & 5.674 & 2.7 & 9.3 & NB & L & 6 \\
13 & LAE J1044$-$0123         & 5.655 & 4.1 & 14  & NB & H & 9 \\
14 & BDF 2:19                 & 5.6488 & 2.5 & 8.7 & SF & L & 4 \\
15 & BR1202$-$0725 LAE & 5.64  & \nodata & \nodata & NB & L & 10 \\
16 & F36246-1511       & 5.631 & \nodata & \nodata & SF & L & 5 \\
17 & HDF 4-473.0       & 5.60  & 1.0 & 3.4 & SF & L & 11 \\
18 & Abell 2218 a\tablenotemark{f} & 5.576 & 6.2\tablenotemark{g}
                                & 0.64 & SF & L/H & 12 \\
19  & 0140+326 RD1      & 5.348 & 3.5 & 11 & SD & L & 13 \\
20 & HDF 3-951.1       & 5.34  & \nodata & \nodata & SF & L & 14 \\
21 & HDF 3-951.2       & 5.34  & \nodata & \nodata & SF & L & 14 \\
22 & J123649.2+621539  & 5.190 & 3.0 & 8.5 & SD & H & 15 \\
23 & Abell 1689\_3             & 5.120 & \nodata & \nodata & SF & L & 16 \\
24 & BDF 1:26                 & 5.0558 & 0.35 & 0.93 & SF & L & 4 \\
25 & BDF 1:18                 & 5.0175 & 0.24 & 0.63 & SF & L & 4 \\
\enddata
\tablenotetext{a}{Observed Ly$\alpha$ flux in units of
$10^{-17}$ ergs s$^{-1}$ cm$^{-2}$.
}
\tablenotetext{b}{Absolute Ly$\alpha$ luminosity in units of
$10^{42}$ ergs s$^{-1}$. If the source is lensed
(i.e., HCM 6A and Abell 2218 a), we give
the unlensed luminosity estimated in the reference.
}
\tablenotetext{c}{Discovery method. NB = imaging survey
with a narrow-passband filter, SF = spectroscopic follow-up in
the field of targeted object, and SD = serendipitous discovery.}
\tablenotetext{d}{All the spectroscopic observations were made using
either LRIS or ESI or both on the W. M. Keck Observatory.
The Echelle mode of ESI provides a high-resolution spectroscopy
(denoted as ^^ ^^ H"; the spectral resolution is higher than
3000) and the low-resolution mode of ESI and LRIS
provide a low-resolution spectroscopy (denoted as ^^ ^^ L";
the spectral resolution is lower than 1000).
}
\tablenotetext{e}{1. Kodaira et al. 2003, 2. Hu et al. 2002, 
3. Cuby et al. 2003, 4. Lehnert
\& Bremer 2003, 5. Dawson et al. 2001,  6. Rhoads et al. 2003, 7. Hu et al.
1999, 8. Ajiki et al. 2002, 9. Taniguchi et al. 2003, 10. Hu et al. 1998, 
11. Weymann et al. 1998,
12. Ellis et al. 2001, 13. Dey et al. 1998, 14. Spinrad et al. 1998,
15. Dawson et al. 2002, and 16. Frye et al. 2002.
}
\tablenotetext{f}{Another Ly$\alpha$ emitter Abell 2218 b
is found at the same redshift. But this is the counter lensed
image of Abell 2218 a.
}
\tablenotetext{g}{In Ellis et al. (2001), the Ly$\alpha$ fluxes obtained
with both Keck LRIS and ESI are given. The Ly$\alpha$ flux
given in this table is their average.
}
\end{deluxetable}
}

%%%%%%%%%%%%% High-z Superwind %%%%%%%%%%%%%%%%%%%%%%%%%%%%%%%%%%%%%%%%%%%%%
\section{SUPERWINDS AT HIGH REDSHIFT}

\subsection{Introduction}

Superwind phenomena are often observed in nearby starburst galaxies
such as M82 (e.g., Ohyama et al. 2003a and references therein;
see for a review Taniguchi et al. 1988; 
Heckman et al. 1990). If bursts of massive star formation
occurred in a forming galaxy and the kinetic energy released from
the collective explosions of supernovae exceeds the gravitational
potential energy of the system, a superwind could  blow 
from this young galaxy. The presence of luminous LAEs at 
high redshift suggests strongly that superwind phenomena are
also common for such forming galaxies. 

Historically, galactic wind models have been applied to the
formation of elliptical galaxies in terms of the monolithic
collapse scenario (e.g., Larson 1974; Arimoto \& Yoshii 1987).
Such monolithic collapse models are now considered 
to be unlikely because hierarchical clustering models (i.e.,
cold dark matter models) provide successful explanations of
the structure formation in the universe (e.g., Peebles 1993).
Nevertheless, it is also known that galactic wind models
appear consistent with many observational properties of 
elliptical galaxies in the local universe. 
This dilemma may be reconciled if we consider the following
situation.
It is not necessarily to presume
that a pregalactic cloud is a first-generation gigantic gas cloud.
If a number of subgalactic gas clouds are assembled into one and then
a starburst occurs in its central region, the physical situation
seems to be nearly the same as that of the monolithic collapse.

In the galactic wind scenario,
the initial starburst occurs at the epoch of galaxy formation
in the galaxy center.
Subsequently, massive stars die and then a large number of
supernovae appear. These supernovae could overlap and then evolve
into a so-called superbubble. If the kinetic energy deposited to the
surrounding gas overcomes the gravitational potential energy of the
galaxy, the gas clouds are blown out into the intergalactic space
as a superwind (e.g., Heckman et al. 1990).
Galactic superwinds are now considered to be one of the key issues for
understanding the interaction and evolution of both galaxies and
intergalactic matter (e.g. Heckman 1999; Taniguchi \& Shioya 2001).
In order to improve our knowledge of galactic superwinds at high
redshift, a large sample of superwind candidates at $z > 3$ is necessary.
Recent optical observations have found some possible candidates
for superwind galaxies at $z > 5$ (Dawson et al. 2002;
Ajiki et al. 2002; see also Frye et al. 2002).
In this section, we give a summary of such observations.

\subsection{J123649.2+621539 at $z=5.19$}

One interesting object among the LAEs given in Table 1
is J123649.2+621539 at $z=5.190$
which was found serendipitously in the HDF-North flanking fields
(Dawson et al. 2002). Its Ly$\alpha$ emission-line profile shows a
sharp blue cutoff and broad red wing emission, both of which are often
observed in star-forming systems with prominent wind outflows. These
features are also expected from radiative transfer in an expanding
envelope. Therefore, Dawson et al. (2002) suggested that the
Ly$\alpha$ profile of J123649.2+621539 is consistent with a superwind
with a velocity of $\sim$ 300 km s$^{-1}$; the Ly$\alpha$ line can be 
decomposed to the following three components, (1) a narrow central emission
component with $FWHM$ = 280 km s$^{-1}$,  (2) a broad emission component 
with $FWHM$ = 560 km s$^{-1}$ redshifted by 320 km s$^{-1}$ from the central
component, and (3) a broad absorption component with $FWHM$ = 800 km s$^{-1}$
blueshifted by 360 km s$^{-1}$ from the central one.

Such superwind activities are expected to be related to the escape
of Lyman continuum photons because the velocity gradient in superwinds
could increase the escape fraction. However, Dawson et al. (2002) obtained 
a very small escape fraction, $f_{\rm esc} \sim 0.1 \pm 0.3$. On the other hand,
Steidel et al. (2001) obtained $f_{\rm esc} \gtrsim 0.5$
for 29 Lyman break galaxies at $<z>=3.40 \pm 0.09$.
In order to improve the measurement of $f_{\rm esc}$ for LAEs at $z>5$,
we need a much larger sample of superwind galaxies at $z>5$.

\subsection{LAE J1044-0130 at $z=5.69$}

During the course of a new search for Ly$\alpha$ emitters at
$z \approx 5.7$ using Subaru Telescope,  Ajiki et al. (2002) found a 
candidate superwind galaxy at $z =
5.69$.  Its optical thumbnails and optical spectrum are shown in 
Figs. 2 and 3, respectively.
Its emission-line shape shows the sharp cutoff at wavelengths
shortward of the line peak and the presence of
the excess red-wing emission seems secure (Fig. 3).  The $FWHM$ of the
Ly$\alpha$ emission is measured to be 340$\pm$110 km s$^{-1}$ and the
full width at zero intensity (FWZI) is estimated to be 890$\pm$110 km
s$^{-1}$. These properties are similar to those of the Ly$\alpha$ emitter
at $z=5.190$, J123649.2+621539, found by Dawson et al. (2002).

%%%%%%%%%%%%%%%%%%%%%%%%%%%%%%%%%%%%%%%%%%%%%%%%%
%   Fig. 2 Thumbnails of LAE 1044-0130
%%%%%%%%%%%%%%%%%%%%%%%%%%%%%%%%%%%%%%%%%%%%%%%%%
\begin{figure}
\epsfysize=6.0cm \epsfbox{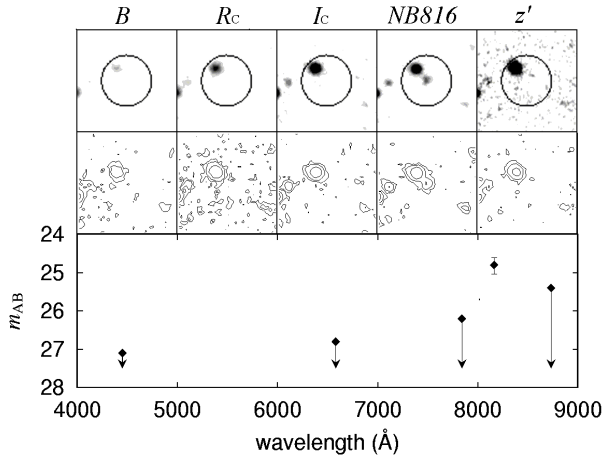}
%\plotone{fig2.ps}
%\vspace{10cm}
\caption[]{
Thumbnail images of LAE J1044-0130 (upper panel)
also displayed as contours (middle panel) taken from Ajiki et al. (2002).
The angular size of the circle
in each panel corresponds to 8$\arcsec$.  The lower panel shows
the spectral energy distribution (in magnitudes). From our optical
spectroscopy, we find that
a galaxy located at 2$\farcs$4 northeast of LAE J1044$-$0130 shows
two emission lines at $\lambda \simeq 6710 - 6720$~\AA.
Since these lines can be identified as the [O
{\sc ii}]$\lambda$3727 doublet redshifted to $z=0.802 \pm 0.002$
(see Figure 9).
This foreground galaxy might enhance
the image of LAE J1044$-$0130 due to gravitational lensing.
However, since there is no counter image of the LAE J1044$-$0130
in our deep NB816
image, the magnification factor might be less than a factor of 2
(Shioya et al. 2002b).
\label{fig2}
}
\end{figure}

%%%%%%%%%%%%%%%%%%%%%%%%%%%%%%%%%%%%%%%%%%%%%%%%%
%   Fig. 3 ESI Spectrum of LAE 1044-0130
%%%%%%%%%%%%%%%%%%%%%%%%%%%%%%%%%%%%%%%%%%%%%%%%%
\begin{figure}
\epsfysize=8.0cm \epsfbox{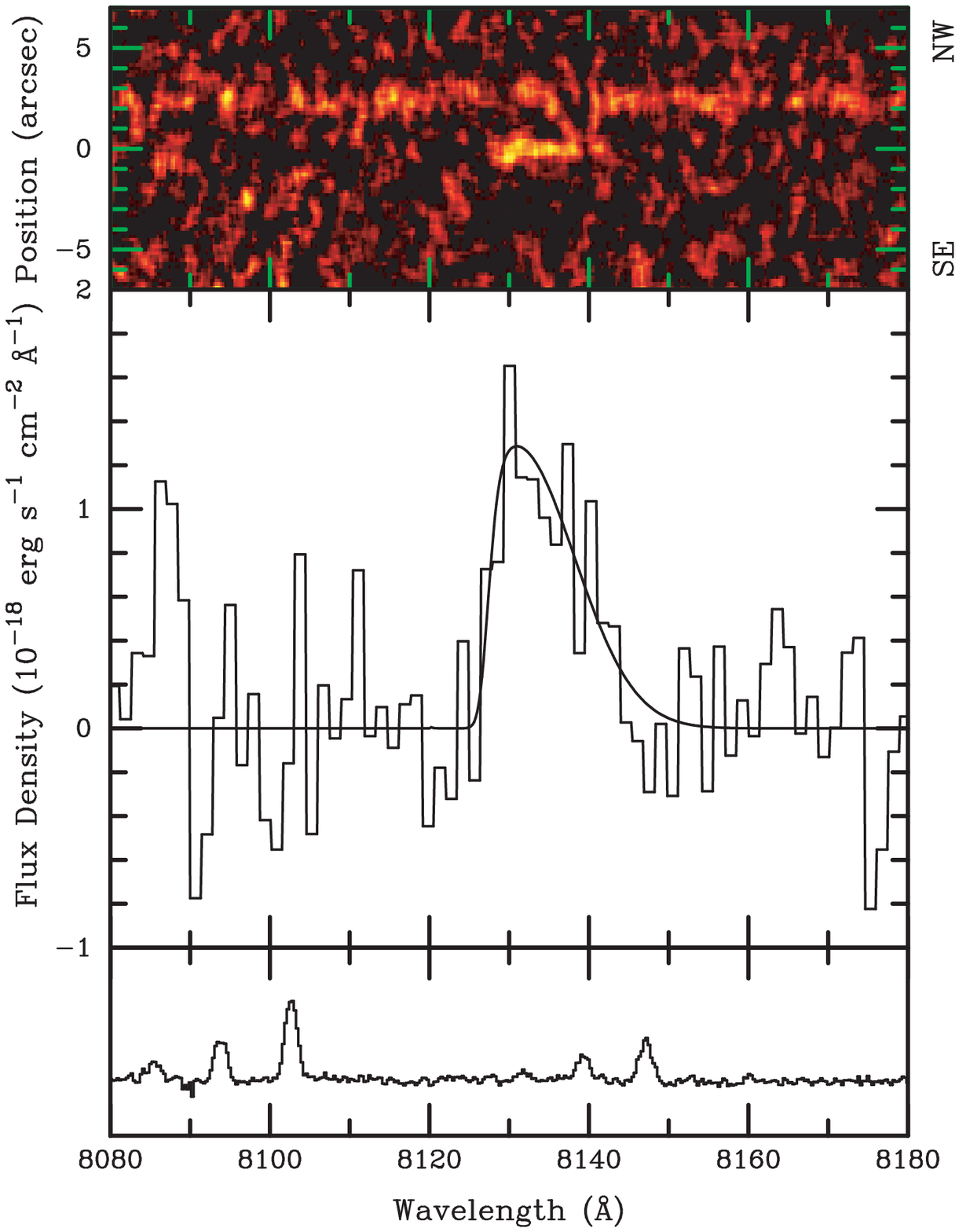}
%\plotone{fig3.ps}
%\vspace{10cm}
\caption[]{
The optical spectrogram (upper panel)
and one-dimensional spectrum (middle panel) of LAE J1044-0130
obtained with ESI on Keck II ($R \sim 3400$) taken from Ajiki et al. (2002).
The model profile fit is
shown by the thick solid curve (see text).  Sky (OH airglow) emission
lines are shown in the lower panel.
\label{fig3}
}
\end{figure}

We note that approximately half of the intrinsic Ly$\alpha$
emission from LAE J1044-0130 could be absorbed by intergalactic atomic
hydrogen (e.g., Dawson et al. 2002). In order to reproduce the
observed Ly$\alpha$ emission-line profile a two-component profile fit
was made using the following assumptions:\ 1) the intrinsic Ly$\alpha$
emission line profile is Gaussian, and 2) the optical depth of the
Ly$\alpha$ absorption increases with decreasing wavelength shortward
of the rest-frame Ly$\alpha$ peak. The resulting fit is shown in
Fig. 3 (thick curve), which corresponds to the following emission
and absorption line parameters. [1] Ly$\alpha$ emission: the line
center, $\lambda_{\rm c, em} = 8031$ \AA, the line flux, $f_{\rm
em} \simeq 2.4 \times 10^{-17}$ ergs s$^{-1}$ cm$^{-2}$, and the line
width, $FWHM_{\rm em} \simeq 650$ km s$^{-1}$; [2] Ly$\alpha$
absorption: the line center, $\lambda_{\rm c, abs} = 8123$ \AA, the
optical depth at the absorption center, $\tau_{\rm abs} \simeq 10$,
and the line width, $FWHM_{\rm abs} \simeq 175$ km s$^{-1}$.  This
analysis suggests that the total Ly$\alpha$ emission-line flux amounts
to $1.7 \times 10^{-17}$ ergs s$^{-1}$ cm$^{-2}$. Then the absorption 
corrected, total Ly$\alpha$
luminosity is estimated as $L$(Ly$\alpha$) $\sim 6.1 \times 10^{42} ~ h_{0.7}^{-2}$
ergs s$^{-1}$, giving the star formation rate,
$SFR \simeq  6  ~h_{0.7}^{-2} ~ M_\odot$ yr$^{-1}$.

\subsection{Lyman$\alpha$ Blobs at $z \approx 3.1$}

Recent narrow-band imaging surveys have revealed the presence of
very extended Ly$\alpha$ emitters at high redshift, $z \sim$ 2 - 3
(Keel et al. 1999; Steidel et al. 2000). 
Three sources found by Keel et al. (1999) are all strong C {\sc iv} emitters
(Pascarelle et al. 1996), and thus 
they are photoionized by the central engine of active galactic
nuclei (AGNs).
On the other hand, since two sources found by Steidel et al. (2000) have no evidence
for the association with AGNs, their origin has been in debate. 
These two sources are called Ly$\alpha$ blobs (LABs).
The observational properties of the LABs found by Steidel et al. (2000) 
are summarized as follows;
1) the observed Ly$\alpha$ luminosities are $\sim 10^{43}$ 
ergs s$^{-1}$, 2) they appear elongated morphologically,
3) their sizes amount to $\sim$ 100  kpc,
4) the observed line widths amount to $\sim 1000$ km s$^{-1}$, 
5) they are not associated with strong radio-continuum sources
such as powerful radio galaxies, and
6) they are strong submm emitters (Chapman et al. 2001). 

As for the origin of LABs, two alternative ideas have been proposed.
One is that these LABs are
superwinds driven by the initial starburst in galaxies because all the above
properties as well as the observed frequency of LABs can be explained
in terms of the superwind model
(Taniguchi \& Shioya 2000). More recently, Taniguchi, Shioya, \& 
Kakazu (2001) found that the observed
spectral energy distribution of one of them, LAB1, found by Steidel et al.
(2000) is quite similar to that of Arp 220 which is a typical ultraluminous
starburst/superwind galaxy in the local universe.
This suggests strongly that the superwind model proposed by
Taniguchi \& Shioya (2000) is applicable to LAB1.
It is remarkable that LAB1 is more luminous in the infrared by a factor of
 20 than Arp 220, and thus Taniguchi et al. (2001) suggest that 
LAB1 is an archetypical example of the hyperwind galaxy.

The other idea is that LABs are cooling radiation from protogalaxies
in dark matter halos (Haiman, Spaans, \& Quataert 2000; Fardal et al. 2001;
see also Fabian et al. 1986; Hu 1992). Standard cold dark matter models
predict that a large number of dark matter halos collapse at high redshift
and they can emit significant Ly$\alpha$ fluxes through collisional excitation
of hydrogen. These Ly$\alpha$ emitting halos are also consistent with
the observed linear sizes, velocity widths, and Ly$\alpha$ fluxes of the
LABs. However, it is uncertain how much far infrared and submillimeter
continuum emission can be emitted because little is known about the dust content
and its spatial distribution in such dark matter halos.
However, this model cannot explain the huge submm luminosity observed in
LAB1 (Chapman et al. 2001). Therefore, the superwind model is much preferred.
Recent optical spectroscopy of LAB1 using Subaru telescope 
also suggested that LAB1 is a superwind galaxy (Ohyama et al. 2003b). 

%%%%%%%%%%%%%%%%%%%%%%%%%%%%%
% Fig. 4 (SED of LAB1)
%%%%%%%%%%%%%%%%%%%%%%%%%%%%%
\begin{figure}
\epsfysize=12.0cm \epsfbox{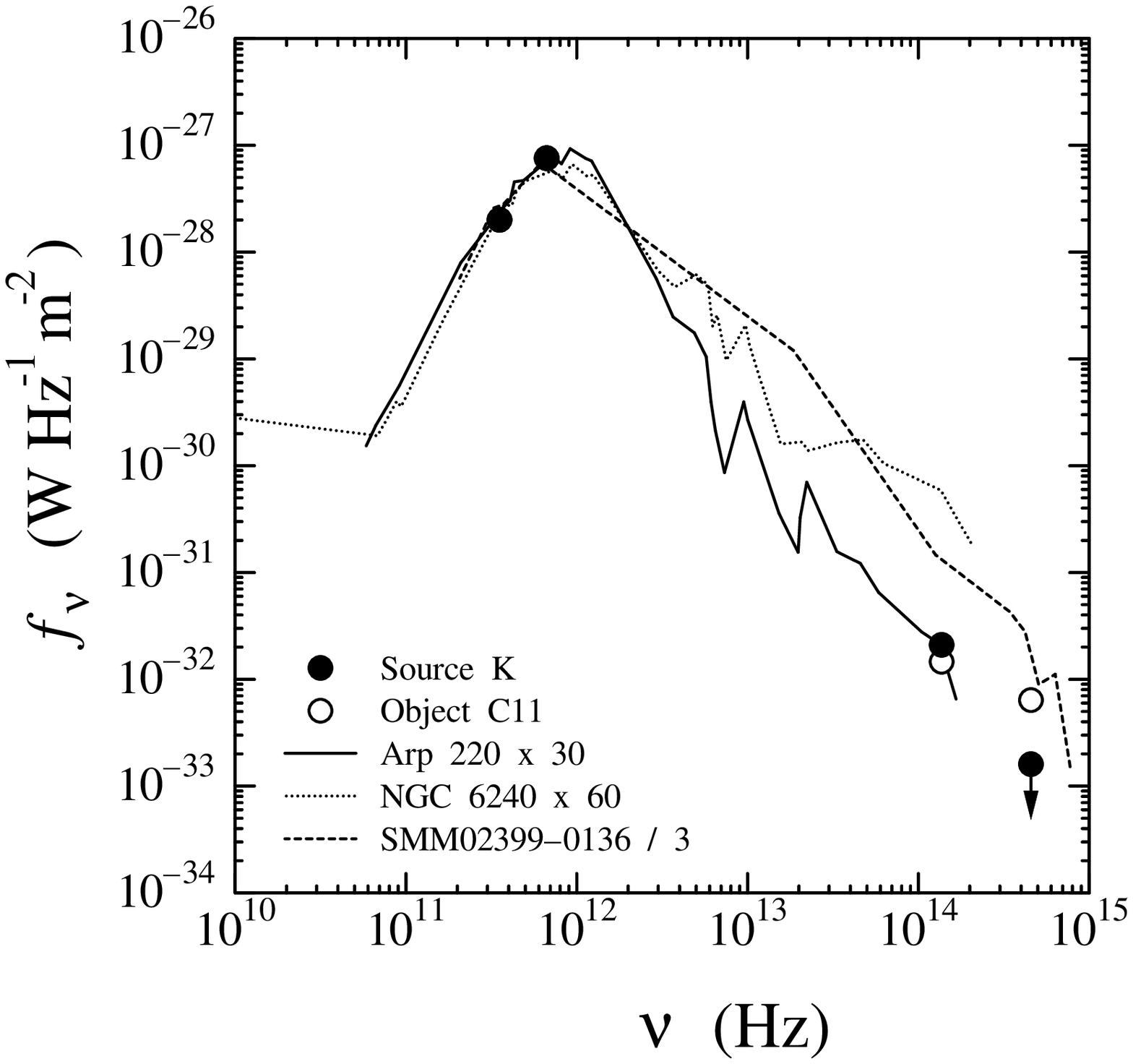}
%\plotone{fig4.eps}
%\vspace{10cm}
\caption[]{
Spectral energy distribution of LAB1 taken from Taniguchi et al. (2001).
The $R$- and $K$-band photometric data of the source K are shown by
filled circles while those of C11 are shown by open circles
(see Steidel et al. 2000).
For comparison,
we show the SEDs of Arp 220 (solid line), NGC 6240 (dotted line),
and SMM 02399$-$0136 (dashed line).
\label{fig4}
}
\end{figure}

In summary, it is likely that LABs are superwinds at $z \sim 3$. 
Another merit of the superwind model is that there seems to be a natural
evolutionary link from
dust-enshrouded (or dusty) submm sources (hereafter DSSs) to LABs
because the central starburst region in a forming elliptical galaxy
could be enshrouded by a lot of gas with dust grains (Taniguchi \& Shioya 2000).
Their scenario is summarized as follows;
Step I: The initial starburst occurs in the center of pregalactic gas cloud.
Step II: This galaxy may be hidden by surrounding gas clouds for
the first several times $10^8$ years (i.e., the DSS phase).
Step III: The superwind blows and thus the DSS phase ceases.
The superwind leads to the formation of extended emission-line regions
around the galaxy (i.e., the LAB phase).
This lasts for a duration of $\sim 1 \times 10^8$ years.
And, Step IV: The galaxy evolves to an ordinary elliptical galaxy
$\sim 10^9$ years after the formation.
This superwind model predicts that the LABs are bright at
rest-frame far-infrared if they are high-redshift, luminous analogues
of nearby superwind galaxies like Arp 220.

\subsection{Chain Galaxies}

One interesting new population of galaxies at high redshift are 
chain galaxies (Cowie et al. 1995; van den Bergh et al. 1996;
see Table 5).
Cowie et al. (1995) found 28 candidates of such chain galaxies
in their deep {\it HST} $I$ band (F814W) images of the
two Hawaii deep survey fields, SSA 13 and SSA 22.
van den Bergh et al. (1996) also identified such chain galaxies
in the Hubble Deep Field-North (Williams et al. 1996);
HDF 2--234 (the tadpole or head-tail galaxy) and
HDF 3--531 (the chain galaxy).
The chain galaxies tend to be very straight in morphology and their average
major-to-minor axial ratio is $\sim$ 5.
Based on these observational properties,
Cowie et al. (1995) proposed that the chain galaxies comprise
a new population of forming galaxies
at high redshift ($z \sim$ 0.5 -- 3)
because they are bluer on average
than galaxies with similar $I$ magnitudes studied by them.
In particular, they are very blue in
$B - K$, suggesting that they are not relatively normal
galaxies in the rest ultraviolet band and that the peculiar
morphologies are not a consequence of the distribution of the
star-forming regions.

%-----------------------------------------------------
%    Table 5
%-----------------------------------------------------
{
\scriptsize
\tablenum{5}
\tablewidth{6.5in}
\begin{deluxetable}{lcc}
\tablecaption{%
Chain galaxies found by Cowie et al. (1995) and van den Bergh et al. (1996)
}
\tablehead{
   \colhead{} &
   \colhead{Cowie et al. (1995)} &
   \colhead{van den Bergh et al (1996)}
}
\startdata
Field & SSA 13 \& SSA 22 & HDF-N \nl
Area  & 10.7 arcmin$^2$ & 5.3  arcmin$^2$ \nl
$N_{\rm chain}(I_{814} \leq 25)$\tablenotemark{a} & 24 & 11 \nl
$N_{\rm chain}(I_{814} > 25)$\tablenotemark{b} & 4 & \nodata \nl
$\sigma_{\rm chain}(I_{814} \leq 25)$\tablenotemark{c} & 2.2 arcmin$^{-2}$ &
        2.1 arcmin$^{-2}$ \nl
$\sigma_{\rm chain}(I_{814} >  25)$\tablenotemark{d} & 0.4 arcmin$^{-2}$ & \nodata \nl
$<a/b>$\tablenotemark{e} & $4.8 \pm 2.1$ & $3.1 \pm 1.3$ \nl
\enddata
\tablenotetext{a}{The observed number of chain galaxies with $I_{814} \leq 25$.}
\tablenotetext{b}{The observed number of chain galaxies with $I_{814} > 25$.}
\tablenotetext{c}{The surface density of chain galaxies with $I_{814} \leq 25$.}
\tablenotetext{d}{The surface density of chain galaxies with $I_{814} > 25$.}
\tablenotetext{e}{The average value of apparent major-to-minor axial ratio.}
\end{deluxetable}
}

As for a possible formation mechanism of chain galaxies,
Taniguchi \& Shioya (2001) proposed the following scenario.
Successive merging of subgalactic gas clumps results in
the formation of a galaxy with a mass of $10^{11-12} M_\odot$
at redshift $z \sim 5$.
Subsequently, supernova explosions occur inside the galaxy
and then blow out as a galactic wind (or a superwind).
This wind expands into the intergalactic space and then causes a
large-scale shell with a radius of several hundreds of kpc.
Since this radius may be smaller than the typical separation between
galaxies, interactions of shells may also occur, resulting in
the formation of a large-scale gaseous slab.
Since the shell or the slab can be regarded as a gaseous sheet,
filament-like gravitational instability is expected to occur.
Further gravitational instability occurs in each filament,
leading to intense star formation along the filament. This is the
chain galaxy phase.
The filament collapses gravitationally into one spheroidal
system like an elliptical galaxy within one dynamical timescale of
the filament ($\sim 10^8$ yr). Therefore, it seems quite difficult
to identify remnants of chain galaxies.
Since forming galaxies with massive stars could often experience 
the superwind phase in their early lives, this idea may remain
as a plausible mechanism.

%%%%%%%%%%%%%%%%%%%%%%%%%%%%%%
%  Fig. 5 (Chain galaxies)
%%%%%%%%%%%%%%%%%%%%%%%%%%%%%%
\begin{figure}
\epsfysize=9.5cm \epsfbox{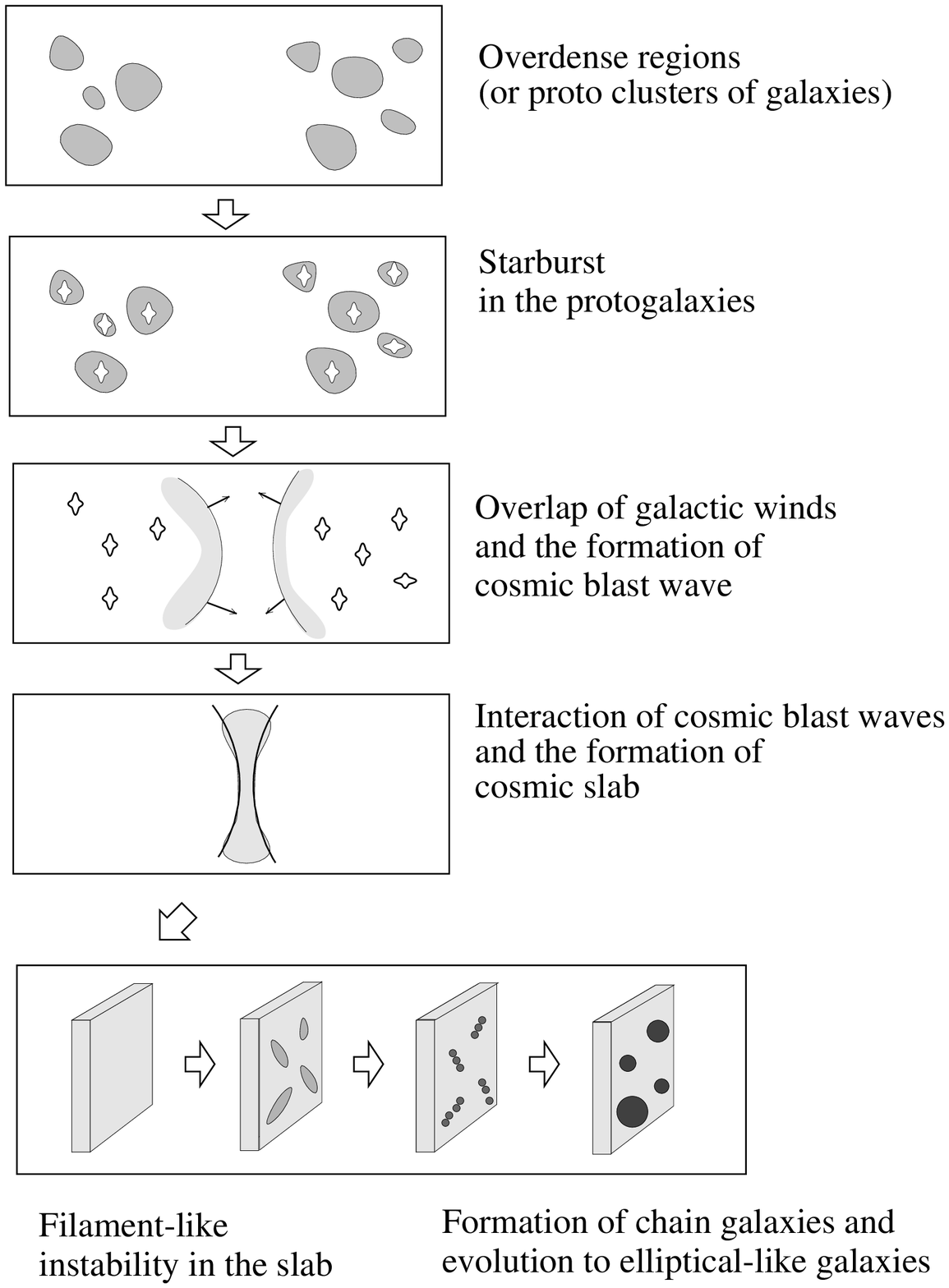}
%\plotone{fig5.ps}
%\vspace{10cm}
\caption[]{
A schematic illustration of the proposed formation mechanism
of chain galaxies taken from Taniguchi \& Shioya (2001).
\label{fig5}
}
\end{figure}

\subsection{Possible Relation to Quasar Absorption Line Systems}

Shocked shells driven by superwinds have been often discussed as
possible agents of some quasar absorption line systems such as
Lyman limit absorption systems (LLSs)
and damped Ly$\alpha$ absorption systems (DLAs)
(Ostriker \& Cowie 1981; Ikeuchi 1981; Chernomordik \& Ozernoy 1983;
Ostriker \& Ikeuchi 1983; Voit 1996).
Indeed, the H {\sc i} column densities of shocked shells driven
by  superwinds are expected to
be $N_{\rm HI} \gtrsim 10^{19}$ cm$^{-2}$ (e.g., Taniguchi \&
Shioya 2001); note that LLSs have $N_{\rm HI} \gtrsim 10^{17}$ cm$^{-2}$
(e.g., Steidel 1990).
If we see such shells from a highly inclined angle (e.g., $\theta_{\rm view}
\gtrsim  80^\circ$) or their tangential sections,
the H {\sc i} column density exceeds $10^{20}$ cm$^{-2}$,
causing damped Ly$\alpha$ absorption systems (DLAs, 
e.g., Peebles 1993; Wolfe et al. 1995).
It is also possible that DLAs may arise
in the last stages of gravitational collapse of the filaments
just as star formation commences.
Small cloudlets breaking into the intergalactic medium (IGM)
may be observed as Ly$\alpha$ forests
with $N_{\rm HI} \lesssim 10^{15}$ cm$^{-2}$.

The superwind scenario may have the following merits.
One is that the projected
distance up to several hundreds kpc is acceptable because 
the absorbing agents are gas clouds in the shocked shells.
The second merit is 
the metal enrichment in the IGM because the superwind
contains a lot of heavy elements.
It has been argued that supernova explosions lead to the chemical
enrichment of the IGM at high redshift (e.g., Ostriker \& Gnedin 1996;
Miralda-Escud\'e \& Rees 1997).
We estimate the metal enrichment due to the superwind.
The mass of metal ejected from a star is $m_Z = \epsilon_Z m_*$
where $\epsilon_Z$ is the mass fraction of metal with respect to
the stellar mass. Thus the total metal enrichment due to the
superwind can be estimated as

\begin{equation}
\Delta Z  ={{\epsilon_Z m_* N_{\rm SN}} \over 
{M_{\rm shell}}} \simeq 4.2 \times 10^{-3} \epsilon_Z
\end{equation}
where we adopt $m_* = 10 M_\odot$ as the typical mass of progenitors of
type II supernovae.
We then obtain $\Delta Z  \simeq  4.2 \times 10^{-4}$ if $\epsilon_Z = 0.1$.
Since  the metal abundance observed in
Ly$\alpha$ forests (e.g., Pettini et al. 1997; Songaila \&
Cowie 1996; Lu et al. 1998) is
of the order of 0.01 $Z_\odot$ where $Z_\odot$ is the solar metal
abundance in mass ($Z_\odot = 0.02$); i.e., $Z_{\rm IGM}
\simeq 2 \times 10^{-4}$. Therefore, the metal
ejection as a result of the superwind from a forming massive galaxy
may be responsible for the observed metal abundance in the IGM
at high redshift.

Spectroscopic evidence for such a shocked-shell absorber
was obtained for two Mg {\sc ii} absorbers at $1<z<2$ (Bond
et al. 2001). These Mg {\sc ii} absorbers show symmetry-inverted 
structure; i.e., the absorption profile has a sharp drop in optical depth 
near the center of the profile and strong black-bottomed absorption
on either side. Future searches for such the symmetry-inverted 
structure in higher redshift absorbers will be recommended.
Since it takes $\sim$ 1 Gyr to develop large-scale superwind-driven
bubbles (e.g., Arimoto \& Yoshii 1987), the presence of
superwind activity gives us information on the formation redshift
of galaxies; e.g., if the formation redshift is
$z_{\rm f} \sim 10$, we may observe superwinds at $z \sim 3$
preferentially. Therefore, any searches for superwinds beyond $z=5$
will provide strong constraints on the formation epoch of massive
star formation beyond $z=10$.

\section{SUBGALACTIC POPULATIONS AT HIGH REDSHIFT}

\subsection{Introduction}

Hierarchical models of structure formation imply that galaxies were
constructed through successive mergers of smaller gaseous clumps
(``building blocks'') in the early universe. This assembly process is
presumed to have taken place from 0.5 to a few Gyrs after the Big
Bang, or over the redshift interval $z \sim$ 30 -- 1.  Therefore, in
order to elucidate the formation and evolution of galaxies, it is
essential to investigate subgalactic objects at high redshifts.
Due to the cosmological redshift, the surface brightness of objects
decreases with increasing redshift
by a factor of $(1+z)^4$, the
so-called Tolman surface brightness dimming (Tolman 1930).
This makes it difficult to estimate the true diameters of any high-$z$
objects and thus it is hard to discover subgalactic small objects
solely by direct imaging observations.
Hence, it is necessary to find objects with small velocity dispersions,
providing firm dynamical evidence for subgalactic masses.
In this section, we briefly present recent observations of such 
subgalactic populations.

\subsection{Abell 2218 a at $z=5.58$}

A group led by R. Ellis conducted a blind spectroscopic survey
of the appropriate critical lines of several well-constrained lensing
clusters, all of which were imaged by HST. During the course
of their spectroscopic observations, they found a very promising 
subgalactic object at $z=5.576$ (see also Table 4).
This object was found as a pair of emission-line objects, but 
their careful lensing model analysis showed that this pair arises
through a gravitationally lensed source at very high magnification;
the magnification factor is as high as $\simeq$ 30.
Its $I$-band magnitude corrected for this amplification is estimated
as $I \simeq 30$ and the physical size is as small as 150 pc.
The star formation rate is estimated as $\sim 0.5 M_\odot$ yr$^{-1}$.
It is also remarkable that its dynamical mass is only $\sim 10^6 M_\odot$,
being comparable to those of globular clusters. 

\subsection{LAE J1044-0123 at $z=5.66$}

The author's group has carried out a very deep optical imaging survey 
for faint Ly$\alpha$
emitters in the field surrounding the quasar SDSSp
J104433.04$-$012502.2 at redshift 5.74 [see section IV (c)].
In addition to LAE J1044$-$0130 (Ajiki et al. 2002), they also found
another single-line
emitter which was identified as a LAE at $z=5.66$ based on their KeckII/ESI
spectrum (Fig. 6). Its thumbnails are shown in Fig. 7.
The most intriguing property of LAE J1044$-$0123 is that the observed
emission-line width (full width at half maximum; FWHM) of redshifted
Ly$\alpha$ is only 2.2 $\pm$ 0.3 \AA. Since the instrumental spectral
resolution is 1.7 $\pm$ 0.1 \AA, the intrinsic width is only 1.4 $\pm$
0.5 \AA, corresponding to $FWHM_{\rm obs} \simeq$ 52 $\pm$ 19 km s$^{-1}$ or a
velocity dispersion 
$\sigma_{\rm obs} = FWHM_{\rm obs}/(2 \sqrt{2 {\rm ln} 2}) \simeq$
22 km s$^{-1}$. This value is comparable to those of luminous globular
clusters (Djorgovski 1995).

%%%%%%%%%%%%%%%%%%%%%%%%%%%%%%
%  Fig. 6 (Sp. of LAE1044-123)
%%%%%%%%%%%%%%%%%%%%%%%%%%%%%%
\begin{figure}
\epsfysize=8.0cm \epsfbox{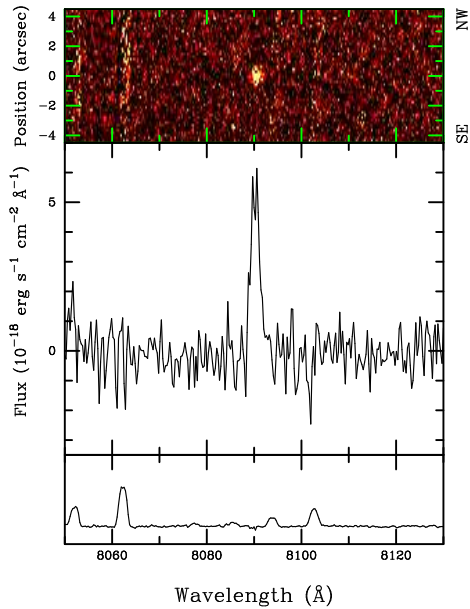}
%\plotone{fig6.ps}
%\vspace{10cm}
\caption[]{
The optical spectrum of LAE J1044-0123 obtained
with the Keck II Echelle Spectrograph and Imager (ESI)
taken from Taniguchi et al. (2003).  We show the
spectrogram in the upper panel and the one-dimensional spectrum
extracted with an aperture of 1.2 arcsec.
\label{fig6}
}
\end{figure}

%%%%%%%%%%%%%%%%%%%%%%%%%%%%%%
%  Fig. 7 (Thumbnail of LAE1044-123)
%%%%%%%%%%%%%%%%%%%%%%%%%%%%%%
\begin{figure}
\epsfysize=6.0cm \epsfbox{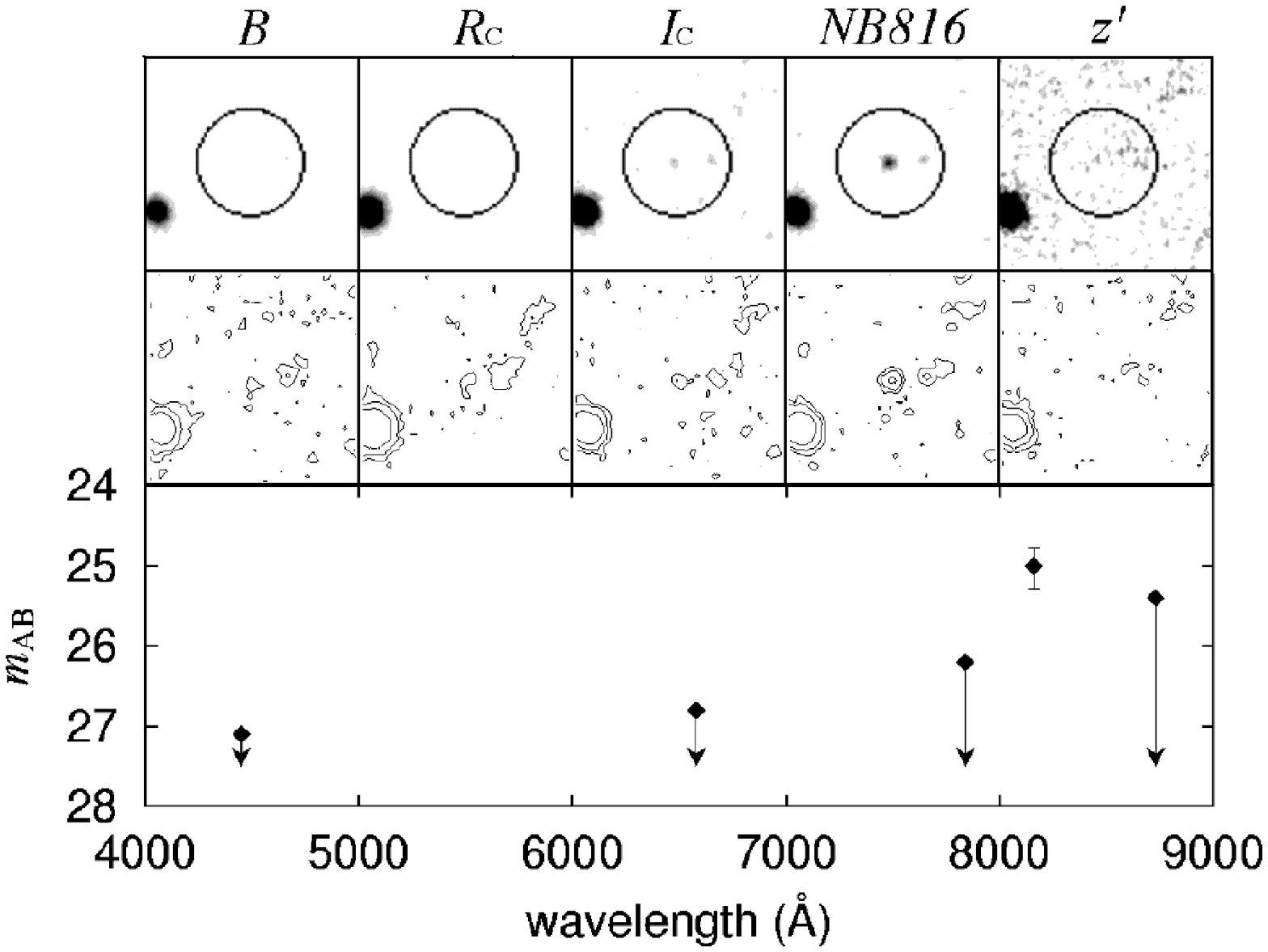}
%\plotone{fig7.ps}
%\vspace{10cm}
\caption[]{
Thumb-nail images of LAE J1044$-$0123 (upper panel)
taken from Taniguchi et al. (2003).
The angular size of the circle in each panel corresponds to 8 arcsec.
Their contours are shown in the middle panel. The lower panel shows
the spectral energy distribution on a magnitude scale.
\label{fig7}
}
\end{figure}

If a source is surrounded by neutral hydrogen, Ly$\alpha$ photons emitted
from the source are heavily scattered. Furthermore, the red damping wing of
the Gunn-Peterson trough could also suppress the Ly$\alpha$ emission line
(Gunn \& Peterson 1965; Miralda-Escud\'e
1998; Miralda-Escud\'e \& Rees 1998; Haiman 2002 and references therein).
If this is the case for LAE J1044$-$0123, we may see only a part of
the Ly$\alpha$ emission. Haiman (2002) estimated that only 8\% of the
Ly$\alpha$ emission is detected in the case of HCM-6A at $z=6.56$
found by Hu et al. (2002).
However, the observed Ly$\alpha$ emission-line profile of LAE J1044$-$0123
shows the sharp cutoff at wavelengths shortward of the line peak.
This property suggests that the H {\sc i} absorption is dominated by
H {\sc i} gas in the system rather than that in the IGM
Therefore, it seems reasonable to assume that the blue half of the
Ly$\alpha$ emission could be absorbed in the case of LAE J1044$-$0123.
Then we estimate a modest velocity dispersion,
$\sigma_0 \sim 2 \sigma_{\rm obs} \sim$ 44 km s$^{-1}$.
Given the diameter of this object probed by the
Ly$\alpha$ emission,  $D \simeq 7.7 h_{0.7}^{-1}$ kpc,
we obtain the dynamical timescale of $\tau_{\rm dyn} \sim
D/\sigma_0 \sim 1.7 \times 10^8$ yr. This would give an upper limit of
the star formation timescale in the system; i.e., $\tau_{\rm SF}
\lesssim \tau_{\rm dyn}$. However, if the observed diameter is
determined by the so-called Str\"omgren sphere photoionized by
a central star cluster, it is not necessary to adopt $\tau_{\rm SF}
\sim \tau_{\rm dyn}$. It seems more appropriate to adopt a shorter
timescale for such high-$z$ star-forming galaxies, e.g.,
$\tau_{\rm SF} \sim 10^7$ yr, as adopted for HCM-6A at $z \approx 6.56$
(Hu et al. 2002) by Haiman (2002).
One may also derive a dynamical mass $M_{\rm dyn} =
(D/2) \sigma_0 ^2 G^{-1} \sim 2 \times 10^9 M_\odot$ (neglecting possible
inclination effects).

At the source redshift, $z=5.655$, the mass of a dark matter halo which
could collapse is estimated as

\begin{equation}
M_{\rm vir} \sim 9 \times 10^6 \left({r_{\rm vir} \over {\rm 1 ~ kpc}}\right)^3
h_{0.7}^{-1} ~ M_\odot
\end{equation}
where $r_{\rm vir}$ is the Virial radius [see equation (24) in Barkana
\& Loeb (2001)]. If we adopt $r_{\rm vir} = D/2$ = 3.85 kpc,
we would obtain $M_{\rm vir} \sim 5 \times 10^8 ~ M_\odot$.
However, the radius of the dark matter halo could be ten times as large as
$D/2$. If this is the case, we obtain $M_{\rm vir} \sim 5 \times 10^{10}
~ M_\odot$ and $\sigma_0 \sim 75$ km s$^{-1}$.
Comparing this velocity dispersion with the observed one, we estimate
that the majority of Ly$\alpha$ emission would be absorbed by neutral
hydrogen.

How massive is this source ?; i.e., $\sim 10^9 M_\odot$ or more
massive than $10^{10} M_\odot$.
If the star formation timescale is as long as the dynamical one,
the stellar mass assembled in LAE J1044$-$0123 at $z=5.655$ exceeds
$10^9 ~ M_\odot$,  being comparable to the
nominal dynamical mass, $M_{\rm dyn} \sim 2 \times 10^9 ~ M_\odot$.
Since it is quite unlikely that most mass is assembled to form stars
in the system, the dark matter halo around LAE J1044$-$0123 would be
more massive by one order of magnitude at least than the above stellar
mass. If this is the case, we could miss the majority of the Ly$\alpha$
emission and the absorption could be attributed to the red damping
wing of neutral hydrogen in the IGM.
Since the redshift of LAE J1044$-$0123 ($z=5.655$) is close to that
of SDSSp J104433.04$-$012502.2 ($z=5.74$), it is possible that
these two objects are located at nearly the same cosmological distance.
The angular separation between LAE J1044$-$0123 and SDSSp
J104433.04$-$012502.2, 113 arcsec, corresponds to the linear
separation of 4.45 $h_{0.7}^{-1}$ Mpc.
The Str\"omgren radius of SDSSp J104433.04$-$012502.2 can be estimated
to be $r_{\rm S} \sim 6.3 (t_{\rm Q}/2\times 10^7 ~ {\rm yr})^{1/3}$
Mpc using equation (1)
in Haiman \& Cen (2002) where $t_{\rm Q}$ is the lifetime of the quasar
(see also Cen \& Haiman 2000).
Even if this quasar is amplified by a factor of 2 by gravitational
lensing (Shioya et al. 2002b), we obtain $r_{\rm S} \sim 4.9$ Mpc.
Therefore, it seems likely that the IGM around LAE J1044$-$0123
may be ionized completely. If this is the case, we cannot expect that
the Ly$\alpha$ emission of LAE J1044$-$0123 is severely absorbed
by the red damping wing emission.
In order to examine which is the case, $L$-band spectroscopy is 
recommended because the redshifted [O {\sc iii}]$\lambda$5007 emission
will be detected at 3.33 $\mu$m, which can be detected by
the James Webb Space Telescope.

\section{On-Going Deep Survey Programs}

\subsection{The LALA Survey}

The LALA (Large Area Lyman Alpha) Survey  has been conducted
by using the wide-field camera (a 36$^\prime \times 36^\prime$ field of view)
on the Kitt Peak National Observatory's 4m Mayall telescope.
Their broad-band images are shared in collaboration with
the NOAO Deep Wide-Field Survey (Januzzi \& Dey 1999).
Together with narrow-band filters, this camera has been
used to search for high-$z$ Ly$\alpha$ emitters at $z \approx 4.5$
(Rhoads et al. 2000;  Malhotra \& Rhoads 2002) and at $z \approx 5.7$
(Rhoads \& Malhotra 2001; Rhoads et al. 2003); see also section II (b).

Their first search for LAEs at $z \approx 4.5$ gives a surface density of
LAEs at $z \approx 4.5$, $1.1 \times 10^4 \pm 700$ deg$^{-2}$
(Rhoads et al 2000). 
Their follow-up optical spectroscopy led to the identification of an
LAE at $z = 4.52$. However, they also identified two low-$z$ interlopers,
suggesting that the success rate of LAE identification is $\sim 1/3$.
A similar success rate is also reported by Kodaira et al. (2003)
for their LAE search at $z \approx 6.6$.
They noted that their LAE candidates at $z \approx 4.5$ tend to have 
larger emission-line equivalent width with respect to the model 
prediction (Malhotra \& Rhoads 2002).
They interpreted this as evidence for massive-star
enhanced star formation in young galaxies. Detailed follow-up spectroscopy
will be necessary to confirm it.

Rhoads \& Malhotra (2001) also made a search for LAEs at $z \approx 5.7$
using the two narrowband filters centered at 8150 \AA ~ (NB815)
and 8230 \AA (NB823),
each of which has FWHM of 75 \AA. They found seven LAE candidates
using the NB815 filter and six using the NB823 one.
Rhoads et al. (2003) made optical follow-up spectroscopy of four LAE
candidates and the found three out of four appear to be LAEs at $z \simeq$ 5.67,
5.70, and 5.75; the remaining one object is too faint to be detected
in their spectroscopy.

\subsection{CADIS}

Another interesting deep survey program is
Calar Alto Deep Imaging Survey (CADIS).
This project uses the 2.2 m and 3.5 m telescopes at
Calar Alto Observatory in Spain and observe 10 sky fields,
each of which has a 120 arcmin$^2$ area (see for a review,
Thommes 1999; see also York et al. 2001, 2003; Hippelein et al. 2003).
Although Thommes et al. (1998) found seven candidates for LAE
at $z \sim 5.7$, their follow-up spectroscopy did not confirm
their identification (see Stern \& Spinrad 1999).

Recently, Maier et al. (2003) reported their new results 
on surveys for LAEs with use of their three narrow 
filter bands centered at 7000 \AA, 8200 \AA, and 9200 \AA;
note that they used an imaging Fabry-Perot interferometer.
They found 5 bright LAE candidates at $z \sim 4.8$ and 
11 ones at $z \sim 5.7$. 

One important characteristic of the CADIS survey is that 
13 intermediate-band (or medium-band) filters are used 
in this survey in addition to the usual four broad-band filters
($B$, $R$, $K$, and $K^\prime$). This filter set provides
better classifications of stars, galaxies, and quasars
at various redshifts (e.g., Wolf et al. 2001, 2003). 
The use of such an intermediate-band filter system
is also planned for the Suprime-Cam on the Subaru Telescope
(Taniguchi 2001).

\subsection{The Subaru Deep Field}

The Subaru telescope team
has officially started a large-scale deep survey program in April 2002;
the Subaru Deep Field (SDF) project.
Several pilot papers related to the SDF project have been already published
(Maihara et al. 2000; Ouchi et al. 2003; Kashikawa et al. 2003;
see also Totani et al. 2001a, 2001b, 2001c).
This project already identified two LAEs at $z \approx 6.6$
(Kodaira et al. 2003); see also section VII (b).
The overall design of the SDF project will be given elsewhere.

\section{THE EARLY COSMIC STAR FORMATION HISTORY}

\subsection{Introduction}

As already noted before, the Lyman break method allowed us to
find a large number of high-$z$ galaxies. Although a small part
of LBGs have been confirmed by spectroscopy, photometric redshifts
are reasonably available for most LBGs. Therefore, the LBG data
are highly useful to probe cosmic star formation history in the early
universe (e.g., Madau et al. 1996; Steidel et al. 1999). 
In particular, Madau et al. (1996) investigated the cosmic star
formation history quantitatively; i.e., the star formation rate
density (SFRD) was investigated as a function of redshift (the so-called
Madau plot). Strictly speaking, we have to take account of the SFRD
contributed by hidden populations such as dusty starburst galaxies
(e.g., Hughes et al. 1998; Barger et al. 1998, 1999). 
Yet, it is shown that the SFRD appears to be constant between
$z \approx  1$ and $z \approx 4$ (e.g., Madau et al. 1998; Steidel
et al. 1999). This constancy appears to extend up to $z \approx 5$
although the SFRD tends to show a slow decline with $z$
(Iwata et al. 2003).

\subsection{Viewed from LAEs}

Here a question arises as; How is the SFRD beyond $z=5$ ?
Since it is quite difficult to find LBGs beyond $z=5$ because
they are basically $I$ dropouts and thus we cannot
estimate their photometric redshifts
solely using $z$ band data. Therefore, LAE survey data are 
highly useful in investigations of the SFRD beyond $z=5$.

Recently, Kodaira et al. (2003) have found 73 LAE candidates in the 
Subaru Deep Field (SDF) based on their very deep imaging with the NB921
filter together with $i^\prime$ and $z^\prime$ data.
They then carried out follow-up optical spectroscopy of nine sources
and discovered two LAEs at $z \approx$6.6.
Since the SDF is a blank field and  the lensing effect is expected to
be small in this field,
they can perform a simple statistical analysis
of star formation activity in the investigated volume.
They obtained the average SFR for the two LAEs;
7.1 $\pm$ 2.0
$h_{0.7}^{-2} ~ M_\odot$ yr$^{-1}$, being comparable to those of LAEs at
$z \simeq$ 5.1 -- 5.8 (e.g., Ajiki et al. 2002).
It should be mentioned that the SFR estimated above is a lower limit
because it is quite likely that a blue half or more of the Ly$\alpha$ emission
may be absorbed by H {\sc i} gas and dust grains in the galaxy itself and by
the intergalactic H {\sc i} gas (Miralda-Escud\'e 1998; Miralda-Escud\'e \&
Rees 1998; Cen \& McDonald 2002).
It is also noted that the SFR based on the Ly$\alpha$ luminosity 
tends to be underestimated
by a few times or more than that based on the UV luminosity (see also
Hu et al. 2002).

Assuming that approximately 22\% (=2/9) of
73 LAE candidates are real LAEs at $z \approx$ 6.5 - 6.6,
they obtained a star formation rate density of $\rho_{\rm SFR} \simeq 5.2 \times 10^{-4}
h_{0.7} ~ M_\odot$ yr$^{-1}$  Mpc$^{-3}$ given 
the survey volume, 202,000 $h_{0.7}^{-3}$ Mpc$^{3}$.
Their estimate can be regarded as a robust and first meaningful
lower limit for the star formation rate density beyond $z=6$.
They also compare this value with previous estimates in a so-called
Madau plot (Fig. 8);
note that all the previous estimates are converted to those in the cosmology
adopted in this review.
As shown in Fig. 8, moderate
star formation activity already occurred in the early universe beyond $z=6$.

%%%%%%%%%%%%%%%%%%%%%%%%%%%%%%%%
%   Fig. 8 (Madau plot)
%%%%%%%%%%%%%%%%%%%%%%%%%%%%%%%%
\begin{figure}
\epsfysize=7.5cm \epsfbox{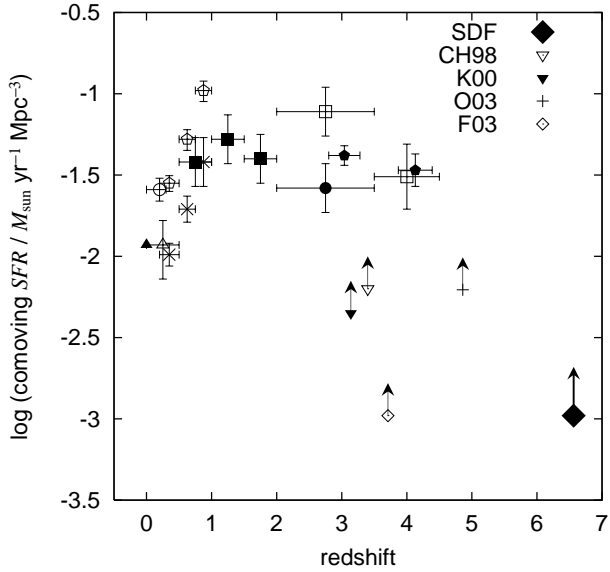}
%\plotone{fig8.eps}
%\vspace{10cm}
\caption[]{
The star formation rate density ($\rho_{\rm SFR}$) as a function of redshift $z$
taken from Kodaira et al. (2003).
Their new estimate at $z \approx 6.6$  (large filled diamond)
is shown together with  the results of previous Ly$\alpha$ searches at $z \sim 3$ - 5
(CH98 = Cowie \& Hu 1998, K00 = Kudritzki et al. 2000, F03 = Fujita et al.
2003, and O03 = Ouchi et al. 2003).
The previous investigations are shown by
filled triangle (Gallego et al. 1996), open triangle (Treyer et al. 1998),
open circle (Tresse \& Maddox 1998), stars (Lilly et al. 1996),
open pentagons (Hammer et al. 1999),
filled squares (Connolly et al. 1997), filled circles (Madau 1998),
and open squares (Pettini et al. 1999).
Other results for Lyman break galaxies between $z$ = 3 - 4
are also shown by filled pentagons (Steidel et al. 1999).
\label{fig8}
}
\end{figure}

Finally, it is also reminded that we do not integrate the SFRDs
of LAE candidates shown in Figure 8, assuming a certain luminosity function from
a lower to a upper limit because there is no
reliable luminosity function for high-$z$ LAEs.
Recently, Ajiki et al. (2003) estimated the Ly$\alpha$ luminosity function
for the LAE samples obtained by both Cowie \& Hu (1998) and their own survey.
The results are shown in Figure 9. If we integrate the SFRDs for both surveys,
we obtain the corrected SFRD shown in Figure 10. The corrected values are quite
similar to those estimated from  Lyman break galaxies at $3 < z < 4$
(see also Bouwens et al. 2003). 
Future careful investigations
will be absolutely necessary to estimate a more reliable contribution
of LAEs to the cosmic SFRD at high redshift.

%%%%%%%%%%%%%%%%%%%%%%%%%%%%%%%%
%   Fig. 9 (LF)
%%%%%%%%%%%%%%%%%%%%%%%%%%%%%%%%
\begin{figure}
\epsfysize=7.5cm \epsfbox{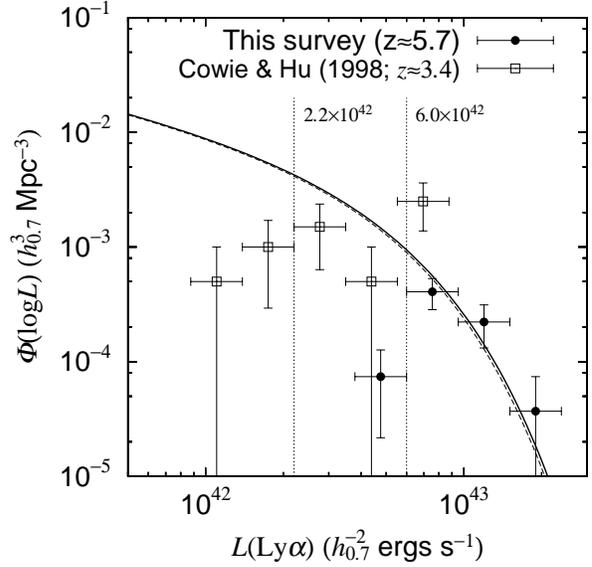}
%\plotone{fig9.eps}
%\vspace{10cm}
\caption[]{
Ly$\alpha$ luminosity function for the two LAE samples obtained by
Ajiki et al. (2003) and Cowie \& Hu (1998). The solid curve is
fro Ajiki et al. (2003) and the dotted one  for Cowie \& Hu (1998).
\label{fig9}
}
\end{figure}

%%%%%%%%%%%%%%%%%%%%%%%%%%%%%%%%
%   Fig. 10 (New Madau plot)
%%%%%%%%%%%%%%%%%%%%%%%%%%%%%%%%
\begin{figure}
\epsfysize=7.5cm \epsfbox{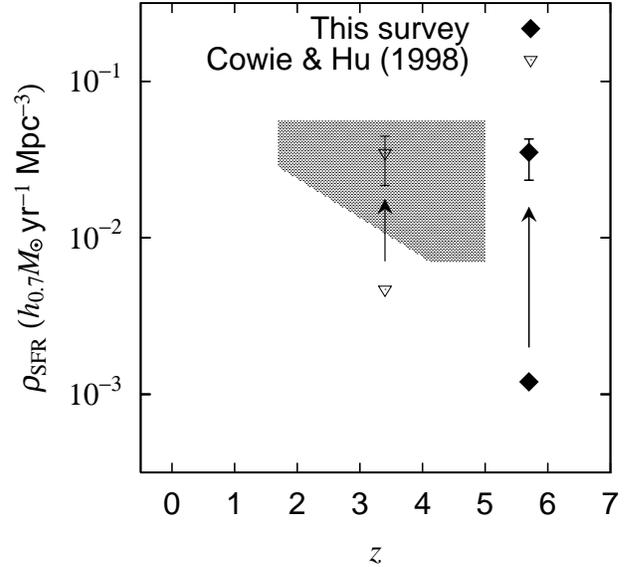}
%\plotone{fig10.eps}
%\vspace{10cm}
\caption[]{
New star formation rate density corrected for the Ly$\alpha$ luminosity
function shown in Figure 9. The filled diamond shows the data point for
Ajiki et al. (2003) and the open inverse triangle is for Cowie \& Hu (1998).
\label{fig10}
}
\end{figure}

\section{COMMENTS ON THE REIONIZATION EPOCH OF INTERGALACTIC MEDIUM}

\subsection{Introduction}

It is believed that our universe began approximately 13 billion years
ago according to the standard big bang cosmology with the adopted
cosmological parameters.
Primeval material (hydrogen, helium, and light elements)
was completely ionized for the first 300,000 years
(its corresponding redshift interval is  $1000 \lesssim z < \infty$)
and then recombined.
Tiny density fluctuations at this stage were thought to evolve
to various cosmological objects like galaxies which are observed
now in the universe (e.g., Peebles 1993; Fukugita, Hogan, \& Peebles 1996;
Bahcall et al. 1999).
The most distant physical information that we can detect is the so-called
cosmic microwave background radiation (CMBR) with temperature of 2.7 K.
This is indeed the redshifted thermal emission from the very young
universe at $z \sim 1000$. Even though we can detect this information,
individual cosmological objects that we can see are distant quasars
and galaxies at $z \sim 6$ (Hu et al.2002 and see Table 4 in this review;
Fan et al. 2003 and references therein).
Were there any light sources  between $z \sim $ 1000 and  $z \sim$ 6 ?
Although we human beings have not yet detected any information from
the universe between $z \sim $ 1000 and  $z \sim$ 6, what we can say now is that
the universe between $z \sim $ 1000 and  $z \sim$ 6 appears dark.
This era is called the dark age of the universe (Rees 1996, 1999).

\subsection{The Ionization State of Intergalactic Matter}

It is evident from the presence of CMBR that the universe once recombined at
$z \sim 1000$. However, it is also known that the intergalactic space
is completely ionized in the universe between $z \sim 6$ and $z \sim 0$.
Its observational evidence is obtained from the so-called Gunn-Peterson
test (Gunn \& Peterson 1965). Since the Gunn-Peterson optical depth at
$z \sim 4$ is much smaller than 1 (Sasaki \& Umemura 1996),
it is strongly suggested that the intergalactic space at this redshift
is really completely ionized (Songaila  et al. 1999; see also Songaila \& 
Cowie 2002).

Another interesting absorption feature found in rest-frame UV spectra of
high-$z$ quasars are Ly$\alpha$ forests. It has been often considered
that these forests are attributed to discrete cloudlets of neutral hydrogen.
However, recent cosmological fluid simulations have suggested that they
are attributed to density fluctuations in the universe
(Cen et al. 1994; Miralda-Escude et al. 1996; Gnedin \& Ostriker 1997;
Zhang et al. 1997). This means that Ly$\alpha$ forests arise from regions
where the fraction of neutral hydrogen is relatively high.
If this is the case, it is not necessarily to distinguish between
Ly$\alpha$ forests and the Gunn-Peterson optical depth. From this reason,
it is now popular to adopt the continuum depression $D_A$ defined
as the integrated continuum absorption between the wavelengths of
Ly$\alpha$ and Ly$\beta$, as a measure of the intergalactic absorption
(Oke \& Korycansky 1982). For example, the most distant quasar known at
$z=6.28$ shows $D_A > 90$\% (Fan et al. 2001).
It is shown that $D_A$ begins to increases rapidly with redshift beyond
$z=3$ and exceeds 0.9 at $z \simeq 5$. Although it was expected that $D_A$
reaches approximately 1 at $z > 5$, recent observations of high-$z$ quasars
 at $z>5$ have shown that $D_A$ appears constant; i.e., $D_A=0.9$.
This led Djorgovski et al. (2001) to suggest that we see the trailing
edge of the reionization at $z \approx 6$, implying that the reionization
occurred at $z > 6$ (Songaila \& Cowie 2002).

\subsection{The Cosmic Reionization}

Recently, Nakamoto, Umemura, \& Susa (2001) made numerical simulations of
the cosmic reionization process by solving the 3-D radiative
transfer equations. This simulation shows that neutral regions begin to
develop at $z \simeq 15$ due to effects of both the self-shielding
and shadowing and then even such regions are subject to ionization
by UV radiation at $z \simeq 9$.
Although the universe appears completely ionized at $z \simeq 7$,
the intergalactic medium is still opaque for ionizing photons.
It is found that the intergalactic medium finally becomes optically
thin for ionizing photons at $z \simeq 5$. The most intriguing result of
this simulation is that the opacity for ionizing photons keeps a high value
even after the completion of reionization of the universe. It is also suggested
that Ly$\alpha$ line emission is subject to strong absorption because
the total absorption cross section for Ly$\alpha$ is similar to
that for photoionization. Accordingly,
the condition of $D_A \approx 1$ does not necessarily mean that
the intergalactic space is neutral.

As mentioned before, there are two alternative reionization sources;
massive stars in galaxies and AGNs like quasars (e.g., Fukugita \& Kawasaki 1994).
The number density of quasars has a significant peak at $z \sim 2$
 and then decreases with increasing redshift. Therefore, recently,
it has been often considered that massive stars in galaxies
are more feasible reionization sources. However, it is still
unknown when and how such galaxies formed in the early universe.
So-called cold dark matter (CDM) models suggest that the first stars
[i.e., Population III (Pop. III) objects] could form at $z \sim 30$.
Recent theoretical considerations suggest that such stars could be
very massive (Abel et al. 1998; Bromm, Coppi, \& Larson 1999; Nakamura \&
Umemura 1999, 2001) although there is an alternative idea that 
intermediate-mass stars could be preferentially formed in zero-metal
pregalactic gas clouds (Yoshii \& Saio 1986; see also Shioya et al. 2002a).

If Pop. III stars are so massive, we have
two alternative ionization sources. One is very hot stars with
temperature of $>10^5$ K and the other is mini AGNs which have hard continuum
 emission. Hot Pop. III stars may have a softer continuum than mini AGNs.
Since the photoionization cross section is proportional approximately to 
$\nu^{-3}$,
harder-energy photons can more easily penetrate into gaseous media, giving rise
to more smoothed ionization structures in the intergalactic space. On the
other hand, the self-shielding effect for softer-energy photons causes
highly-contrasted ionization structures. In either case, these sources could
reionize the universe up to $z \simeq 10$ (Loeb \& Barkana 2001).
This means that the universe was reionized by objects formed in
the dark age and then the reionization process ended in the dark age.

It is considered that Pop. III stars formed in subgalactic clumps at
$z \sim$ 10 -- 30 (Gnedin \& Ostriker 1997).
If these stars could contribute to the reionization, the star formation
density may be $\sim 0.2 M_\odot$ yr$^{-1}$  Mpc$^{-3}$,
being comparable to that at $z \simeq $ 1 -- 4 (Barkana \& Loeb 2000;
Blain et al. 1999). Although massive stars in these subgalactic clumps at
$z \sim$ 10 -- 30 could be feasible reionization sources,
they are too faint to be detected by using any 8m-class telescopes; if we
try to detect $\sim 0.1 L^*$ galaxies using 8-10m class telescopes
(i.e., $J$ dropouts), we need more than 100 hours integration for both
$J$ and $H$ bands (F. Iwamuro, private communication).

It is considered that galaxies with an ordinary mass could form at
$z \sim 10$. Indeed, very high-$z$ galaxies were already found at
$z=6.56$ (Hu et al. 2002) and at $z=6.58$ (Kodaira et al. 2003).
Although new high-$z$ quasars will be found in the existing and future surveys,
finding high-$z$ galaxies seems more important because the
number density of galaxies would be much higher than that of quasars.
In order to understand the reionization process, new systematic searches
for Ly$\alpha$ emitting galaxies are very important although 
the WMAP data suggest that the cosmic reionization epoch may be
$z_{\rm rec} \approx 17 \pm 4$ (Spergel et al. 2003).

\subsection{Further Comments}

Recent discovery of quasars at $z \sim 6$ by the Sloan Digital Sky Survey
(SDSS) made it possible to examine a so-called Gunn-Peterson trough in the
UV spectra of such SDSS high-$z$ quasars (e.g., Fan et al. 2001, 2003;
Becker et al. 2001). Since such new information can be used to estimate
the epoch of the cosmic reionization, deep and wide-field searches for LAEs
beyond $z=6$ gains high importance (see for a review, Loeb \& Barkana 2001).
It seems important that 
there is no continuum emission shortward of the Ly$\alpha$ line
of HCM 6A at $z=6.56$ (Hu et al. 2002). The two LAEs at $z=6.6$
found in Kodaira et al. (2003) also share the same property.
It is known that the radiation damping wing from the neutral H {\sc i}
gas can absorb radiation 20 \AA ~ redward of the Ly$\alpha$
wavelength in the observed frame of the object (Miralda-Escud\'e 1998;
Miralda-Escud\'e \& Rees 1998). Therefore, the presence
of LAEs beyond $z=6$ suggests that the reionization epoch may be earlier
than $z=6$ (e.g., Hu et al. 2002).
However, if a LAE is surrounded by
a cosmological H {\sc ii} region that is made by the LAE itself,
the Ly$\alpha$ emission line can remain observable even if most
intergalactic medium is neutral (Haiman 2002).
In order to pursue this issue, we need a larger sample of LAEs
between $z =6$ and $z=7$ scattered in a large sky area.

\section{NEAR FUTURE PROSPECTS}

\subsection{Introduction}

As demonstrated in this review, more than twenty 
LAEs beyond $z=5$ have been already discovered. It is also
expected that a large number of such LAEs will be found
soon because there are a number of on-going deep survey programs
aiming at finding them. In fact, our group has found more than
20 LAE candidates at $z \approx 5.7$ (Ajiki et al. 2002).
The SDF team has also found $\sim$ 70 LAE candidates at
$z \approx 6.6$. Therefore, we will be able to investigate
the cosmic star formation history probed by a large number of
high-$z$ LAEs soon (perhaps, within a couple of years). 

\subsection{Optical Follow-up Spectroscopy}

The progress in LAE searches beyond $z=5$ may depend on 
the efficiency of follow-up optical spectroscopy.
Although multi-object spectroscopy mode has been already 
ready in most spectrographs on the 8-10m telescopes,
high-spectral resolution (e.g., $R >$ 2000 - 3000) is 
absolutely necessary
to confirm the blue-sharp-cutoff profile which is the most
important property of the Ly$\alpha$ emission line.
The reason for this is that only Ly$\alpha$ can be found
in optical spectra of LAEs beyond $z=5$. Indeed, it has been
often discussed that a single emission line in optical spectra
of faint galaxies can be either Ly$\alpha$ or [O {\sc ii}]$\lambda$3727
(e.g., Stern et al. 2000a and references therein).

Such a single line with large equivalent width (e.g., $>$ 200 \AA)
has been considered to be a probable LAE. However, intense star-forming
galaxies at intermediate redshift also sometimes have such large
equivalent widths of [O {\sc ii}] emission (e.g., Ohyama et al. 1999;
Stern et al. 2000a).
Therefore, single-line objects with a large equivalent width are not
always LAEs at high redshift. The continuum break at wavelengths
shortward of Ly$\alpha$ also provides firm evidence for a LAE
(e.g., Hu et al. 2002). However, the star formation timescale in
LAEs beyond $z=5$ may not be so long (e.g., less than 1 Gyr).
Therefore, LAEs are not always  detectable in the continuum
when their wavelength is 
longer than Ly$\alpha$ that can be probed at $z$ photometric band.
Accordingly, the line shape is the best discriminator of LAEs
from low-$z$ interlopers.
Indeed, if the spectral resolution is high enough to resolve the [O {\sc ii}]
doublet line (see Fig. 11), it is easily to reject low-$z$ interlopers
from high-$z$ LAE candidates. 

%%%%%%%%%%%%%%%%%%%%%%%%
%   Fig. 11 ([OII])
%%%%%%%%%%%%%%%%%%%%%%%%
\begin{figure}
\epsfysize=7cm \epsfbox{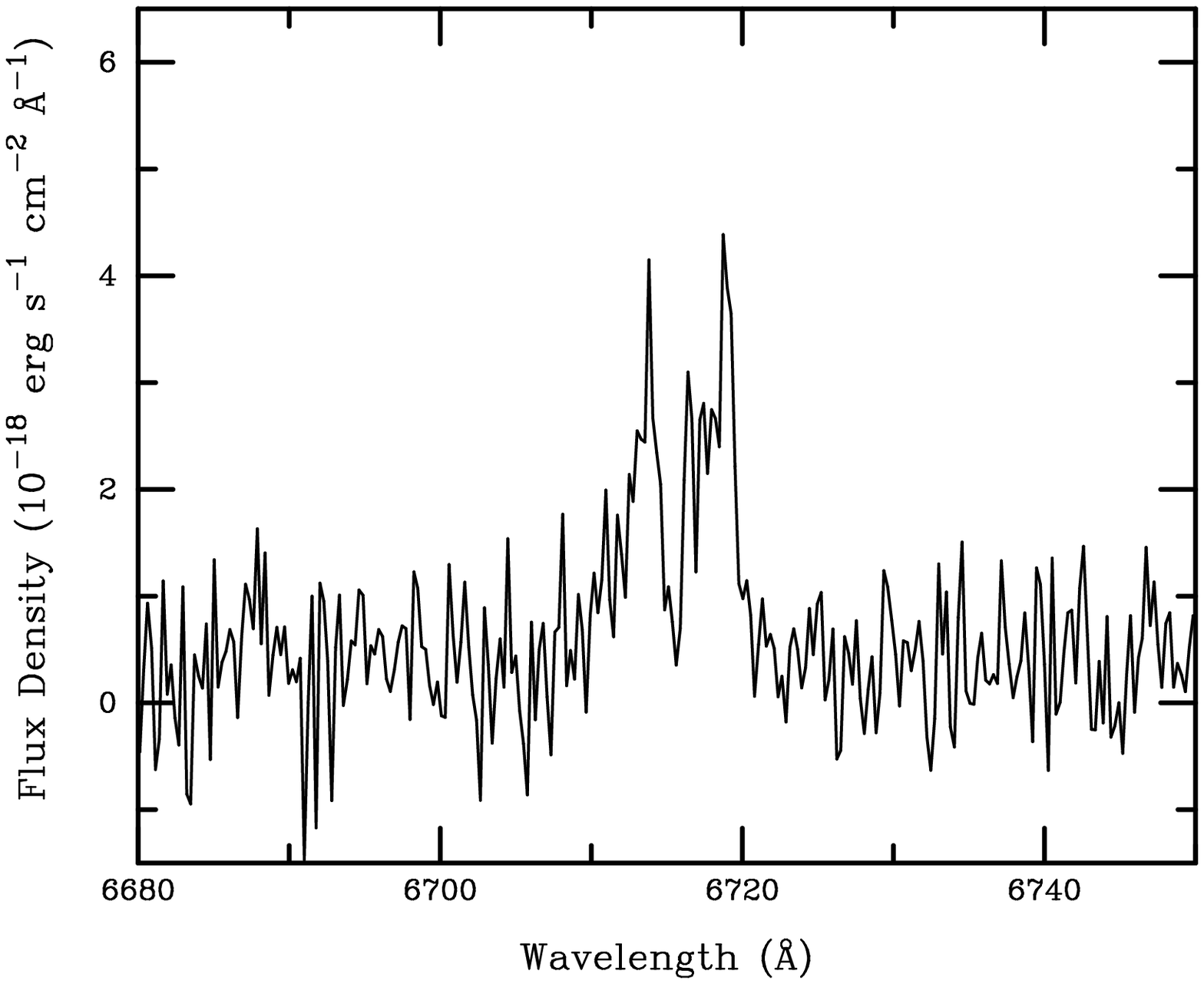}
%\plotone{fig11.ps}
%\vspace{10cm}
\caption[]{
An optical spectrum of the foreground neighbor galaxy ($z=0.802$) of
LAE J1044$-$0130 (Ajiki et al. 2002) obtained with ESI on Keck II
(see Figure 2).
The [O {\sc ii}]$\lambda$3727 doublet is clearly seen.
\label{fig11}
}
\end{figure}

\subsection{New Surveys with Intermediate-Band Filters}

New-type filters, intermediate-band filters, have been recently
introduced in optical deep surveys such as CADIS 
[see section VI (b)]. The MAHOROBA project planned to be made
with use of the Suprime-Cam on the Subaru Telescope also uses
such an intermediate-band filter system. This filter system
has a spectroscopic resolution of $R = 23$ covering wavelengths 
between 3800 \AA ~ and 9900 \AA ~ (Taniguchi 2001); see Fig. 12.

\begin{figure*}
\epsfysize=5.5cm \epsfbox{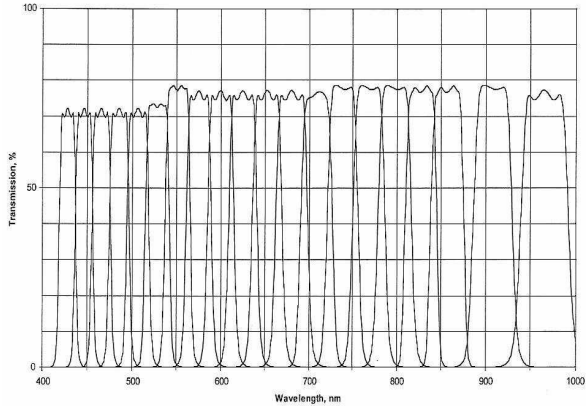}
%\epsscale{1.5}
%\plotone{fig12.ps}
\caption[]{
Transmission curves for the Subaru intermediate-band filters.
Note that the transmission curves of the first two filters are
not drawn in this Figure.
\label{fig12}
}
\end{figure*}

A pilot survey for LAEs at $z \approx 3.7$ is reported by
Fujita et al. (2003), in which one of the intermediate-band filters,
IA574 (centered at 5740 \AA ~ with FWHM of 280 \AA), is used.
Although their survey depth is not so deep, they have found 
six LAE candidates among 23 strong emission-line sources.
Although the use of such an intermediate-band filter is not
useful in searching for very faint LAEs,
a large number of strong LAEs should detected,
because each filter can probe a volume of 
$\sim 10^6$ Mpc$^3$ thanks to the wide-filed coverage of 
the Suprime-Cam. 

\subsection{NIR Surveys}

If we use optical imagers, we cannot find objects
beyond $z=7$ because the Ly$\alpha$ line passes out of the optical
window for such a very high redshift. Therefore, it will become
much more important to promote near-infrared (NIR) deep surveys
(e.g., Pahre \& Djorgovski 1995; see also for papers on broad-band
NIR surveys, Dickinson et al. 2000; Yahata et al. 2000).
One technical problem is that any existing NIR cameras 
do not have a wide-field coverage. The next-generation NIR camera,
MOIRCS, installed on the Subaru telescope may be the widest NIR 
camera with a field coverage is $4^\prime \times 7^\prime$;
note that this may be the widest but its field coverage is 
almost comparable to typical optical imagers on the 8-10m class
telescope like LRIS on Keck I. We need a much wider NIR camera
in the future.

\vspace{0.5cm}

We would like to thank the organizers of this exciting and fruitful
workshop at Pohang, Korea, in particular, Hyung Mok Lee,
and all participants, in particular, 
Renyue Cen, Paul Shapiro, Dave Sanders, Peter Quinn, and Nick Scoville.
We would also like to thank Alan Stockton, Esther Hu, Masayuki Umemura,
Zoltan Haiman, Fumihide Iwamuro, Youichi Ohyama, Christian Maier, 
and Satoru Ikeuchi for their kind communications.
We also thank all members of the SDF project, in particular, K. Kodaira,
H. Karoji, H. Ando, M. Iye, N. Kashikawa, S. Okamura, K. Shimasaku,
\& M. Ouchi.
This work was financially supported in part by
the Ministry of Education, Culture, Sports, Science, and Technology
(Nos. 10044052, and 10304013).

%-------------------------------------------------------------------------

\end{document}